\documentclass[pdflatex,sn-mathphys-num]{sn-jnl}

\usepackage{multirow}
\usepackage{amsmath,amssymb,amsfonts}
\usepackage{mathrsfs}
\usepackage{xcolor}
\usepackage{textcomp}
\usepackage{booktabs}
\usepackage{algorithm}
\usepackage{algorithmicx}
\usepackage{algpseudocode}
\usepackage[section]{placeins}
\usepackage{subcaption}

\begin{document}

\title[Article Title]{Adversarial SQL Injection Generation with  LLM-Based Architectures}

\author[1]{\fnm{Ali} \sur{Karakoc}}\email{ali.karakoc@std.bogazici.edu.tr}

\author*[1]{\fnm{H. Birkan} \sur{Yilmaz}}\email{birkan.yilmaz@bogazici.edu.tr}

\affil[1]{\orgdiv{Department of Computer Engineering, NETLAB}, \orgname{Bogazici University}, \orgaddress{\city{Istanbul}, \country{Turkey}}}

\abstract{SQL injection (SQLi) attacks are still one of the serious attacks ranked in the Open Worldwide Application Security Project (OWASP) Top 10 threats. Today, with advances in Artificial Intelligence (AI), especially in Large Language Models (LLMs), an opportunity has been created for automating adversarial attack tests to measure the defense mechanisms. In this paper, we aim to create a comprehensive evaluation of use cases that utilize LLMs for adversarial SQL injection generation. We introduce two novel LLM-based systems, Retrieval Augmented Generation for Adversarial SQLi (RADAGAS) and Reflective Chain-of-Thought SQLi (RefleXQLi), and compare them with existing baselines against 10 Web Application Firewalls (WAFs) and one execution-based MySQL validator. To perform a comprehensive test, we used six rule-based open-source WAFs (ModSecurity PL1--3, Coraza PL1--3), 2 AI/ML-based WAFs (WAF Brain, CNN-WAF), and 2 commercial WAFs (AWS WAF and Cloudflare WAF). For the LLM models, we used GPT-4o, Claude 3.7 Sonnet, and DeepSeek R1.
Our tests consist of 240 experiments that generate 240,000 payloads and perform 2.2 million tests against WAFs. Our comprehensive evaluation reveals that RADAGAS-GPT4o outperforms other baseline models with a 22.73\% bypass rate. The proposed RADAGAS variants are highly successful on AI/ML-based WAFs (92.49\% on WAF-Brain by RADAGAS-DeepSeek, 80.48\% on CNN-WAF by RADAGAS-Claude), but struggle to bypass rule-based WAFs (0--5.70\% on ModSecurity and Coraza). In addition to these findings, another observation is that creating less diverse payloads achieves more bypasses, however they show poor results if the initially chosen payload is not successful. We observe that our findings provide a comprehensive view on using LLM-based approaches in security testing.} 

\keywords{SQL injection, Large language models, Web application firewalls, Adversarial testing, RAG-based generation, Security benchmarking}

\maketitle

\section{Introduction}\label{sec:introduction}

Web applications are the fundamental part of the modern digital world, and connect people and businesses in different critical domains like healthcare, finance, social networking, e-commerce and government services. Although it is one of the very known and initial web application attacks, SQL Injection vulnerabilities are still one of the major problems persisting on the Open Worldwide Application Security Project (OWASP)'s Top 10 most critical web application security risks~\cite{owasp2021}. The impact of SQL injection attacks can be leveraged with more sophisticated techniques to steal critical information, obtain remote services or denial of services. To protect the web application services against SQL Injection attacks, using advanced defense mechanisms is crucial. However, defense mechanisms do not work with simple installation; there are necessary configuration settings, fine tunings and manual or automatic rule creation that must be done to establish more robust and secure systems. This can be done by adversarial assessment on the systems to see the potential weaknesses.

Traditional approaches to test web applications against SQL injection (SQLi) attacks rely on two main methodologies: manual penetration testing performed by real security experts and automated scanning via mainstream rule-based SQLi automation tools such as SQLmap~\cite{sqlmap2006}. While these methods have proven success revealing SQLi vulnerabilities, they face significant limitations against modern AI/ML based and commercial WAFs. Manual testing is time consuming, expensive and the success totally depends on the expertise of the security tester. Automated rule-based testing is not able to generate novel SQLi attacks to bypass next generation systems and they can be easily recognized by signature based defense systems. In addition to these, both approaches do not perform coverage tests for different variants of the SQLi attacks. The purpose of adversarial testing against defense systems is increasing the defense bar proactively even before an incident happens.

The recent studies in Large Language Models (LLMs) have sparked significant interest in implementation of these approaches to security testing. Models such as GPT-4o~\cite{openai2024gpt4}, Claude Sonnet~\cite{anthropic2024claude}, and DeepSeek~\cite{deepseek2024coder} demonstrated advanced capabilities of understanding and generating code-like structures including SQL syntax. Their ability of understanding syntax and making advanced reasoning positions them as a promising tool for generating adversarial SQL injection payloads. However, the existing work on adversarial testing with LLMs has several challenges: single-model focus, limited parameter exploration, insufficient diversity and feedback mechanisms.

\textbf{Single-Model Focus:} Previous researches mainly evaluate the capability of a single LLM, that limits understanding the behavior of cross model setups and optimal parameter configurations. This prevents understanding the nature of LLM based adversarial SQLi generation comprehensively.

\textbf{Limited Parameter Exploration:} Existing studies demonstrate the performance of adversarial generations under fixed or minimally varied hyperparameter settings. This prevents exploring model specific optimal parameter settings and understanding of their effectiveness.

\textbf{Insufficient Diversity and Feedback Mechanisms:} Current LLM based approaches generally do not put diversity enforcements that result in repetitive SQLi payloads and limit the attack coverage~\cite{Manes2019FuzzingSurvey}. This makes the SQLi generator easily detectable by advanced defense systems~\cite{modsecurity2024}.

To address these challenges and to analyze adversarial SQLi generation with LLMs, we introduce two novel architectures and a comprehensive benchmarking study against modern WAF defenses. Our contributions can be listed as follows:

\begin{enumerate}
\item \textbf{Comprehensive Benchmarking Framework:} We present a systematic evaluation framework to compare LLM SQLi generator models against ten  comprehensive WAFs: AI/ML-based (WAF-Brain, CNN-WAF), rule based open source (ModSecurity PL1-3, Coraza PL1-3), commercial (Cloudflare, AWS WAF), and MySQL execution validation. We created 240 different experiment cases then generated 240,000 SQLi payloads and performed 2.2 million tests against ten WAFs and one execution test that provides an extensive benchmarking on LLM based adversarial SQLi generation.

\item \textbf{Retrieval-Augmented Generation for Adversarial SQLi (RADAGAS):} We introduce \textbf{RADAGAS} architecture that utilizes Retrieval Augmented Generation (RAG) and LLM generation with multi stage diversity and execution filtering. RADAGAS collects and retrieves proven successful attack patterns from curated knowledgebase (OWASP, Portswigger, Github) and generates adversarial SQLi payloads via RAG prompting. We experimented RADAGAS with different foundation models (GPT-4o, Claude 3.7 Sonnet and Deepseek-r1) to demonstrate its cross-model performance.

\item \textbf{Reflective Chain-of-Thought SQLi (RefleXQLi):} Our second proposed method is RefleXQLi that employs four step chain of thought reasoning (WAF Analysis, Strategy Formulation, Payload Design, Refinement) to generate high quality adversarial SQLi payloads followed by dual-LLM validation. The architecture consists of two modules: generator LLM creates SQLi payloads and discriminator LLM refines them iteratively that resulted in generation of payloads with 100\% uniqueness.

\item The experiments demonstrated that RADAGAS variations (22.73\%, 22.09\%, 21.73\%) and RefleXQLi (21.21\%) outperformed existing traditional methods (15.01\%), vanilla zero-shot LLM generator (12.90\%) and GenSQLi (20.35\%)~\cite{babaey2025gensqli} in terms of average WAF bypass rate, showing that both proposed methods are superior to the traditional approaches.

\item  The experiments demonstrated that the temperature settings are model specific and counter intuitive~\cite{Dong2024Survey}. While GPT-4o performed better in lower temperature (T=0.1, 22.73\%), Deepseek-r1 performed better in high temperature (T=0.9, 22.09\%) with more creativity and Claude 3.7 Sonnet has the optimal performance in medium temperature (T=0.6, 21.73\%). This demonstrates LLMs have different characteristics under different temperature conditions in terms of exploitation versus exploration in adversarial generation.

\item \textbf{MySQL Validation as Ground Truth:}
As an additional validation layer, we introduced a vulnerable web application backed by a MySQL server to validate the generated SQL injection payloads. The experiment results show that RADAGAS systems achieve 60-64\% execution success, while the baseline systems stay at 21-37\% execution success against MySQL validation.

\item \textbf{WAF Performance Characterization:}
We demonstrated system specific performance across all 10 WAFs and one MySQL validation layer. RADAGAS-DeepSeek performed high bypass rates on ML based WAFs with 92.49\% on WAF-Brain, 78.39\% on CNN-WAF, GenSQLi achieved the highest rule-based WAF bypass with 16.26\% on ModSecurity PL1, and RADAGAS-GPT4o achieved the highest bypass rate on Cloudflare with 49.50\%. This heterogeneous result set shows that there is no single system that outperforms universally, and the practical implementations should combine multiple approaches together.

\item \textbf{RefleXQLi Chain of Thought Evaluation:} We introduced RefleXQLi which is one of the first implementations of Chain of Thought reasoning combined with dual-LLM architecture for adversarial attack generation. RefleXQLi demonstrated 21.21\% bypass rate with low variance $\sigma=0.37$, that shows explicit reasoning provides more stable output for adversarial payload generation. However, 27.80\% MySQL execution success is lower than RADAGAS's 60-64\% revealing a tradeoff between creative adversarial generation and semantic correctness of the generated outputs.
\end{enumerate}

\section{Related Work}
\label{sec:related}

We organized related work section into three categories: Traditional SQLi generation, LLM based security testing, and diversity in adversarial generation.

\subsection{Traditional SQL Injection Generation}

SQLi vulnerability is known for decades and SQLi testing tools and approaches are quite comprehensive and contain decades of expert knowledge embedded into rule based attack patterns. SQLMap~\cite{sqlmap2006} is one of the industry standard popular tools and extensively used in penetration testing to cover SQLi scenarios since 2006. The tool consists of thousands of hand crafted SQL injection payloads in its database, covering union based, error based, time based blind, boolean based blind and stacked queries across multiple database platforms like MySQL, PostgreSQL, Oracle and MSSQL. SQLMap utilizes automated parameter detection, DBMS fingerprinting and  enumeration strategies to achieve comprehensive coverage. Surveys~\cite{Kindy2011SQLi,Clarke2012SQLBook} state the evolution of attack types and their countermeasures of traditional SQLi frameworks over the past two decades.

Although traditional tools have extensive coverage in testing SQLi vulnerabilities, there are several limitations against modern defense systems. Since the tools generate deterministic payloads, those patterns can be captured easily by signature based WAFs~\cite{appelt2015websec}. ML based detection systems are quite effective on template based approaches since these approaches have weakness on contextual adaptation~\cite{CNNWAF2019,WAFBrain2018} and ML based WAFs can learn the attack characteristics easily. In~\cite{appelt2015websec}, Appelt et al. show that the commercial WAFs are significantly effective on detecting SQLMap generated attack payloads with 94\% prevention rate. Even sophisticated manual tests are ineffective against multi layer defense in depth systems.

\subsection{LLM-Based Security Testing}

SQLiGPT was one of the first uses of LLMs for adversarial security testing~\cite{sqligpt}. The tool examines the capability of GPT model when it generates and detects SQLi payloads with prompt engineering but in a limited scale. In~\cite{llmsqli}, Yang et al. developed a system named LLMSqli to detect SQLi attacks using generalization capabilities of LLMs. Fang et al. introduced an agent based LLM architecture to learn exploiting web vulnerabilities via adaptive feedback loops~\cite{fang2025llm4vuln}. In another approach, Deng et al. designed pentestGPT to perform automated penetration testing~\cite{deng2024pentestgpt}. In~\cite{liu2025promptinject}, Liu et al. introduced a mechanism to detect adversarial prompt injection techniques against LLM based systems.

Babaey and Ravindran introduced a comprehensive framework named GenSQLi to generate adversarial SQLi attacks and generate corresponding WAF rule to create defense for these payloads~\cite{babaey2025gensqli}. GenSQLi employs GPT-4o~\cite{openai2024gpt4} with in context learning. They also evaluated Google Gemini Pro which performed lower bypass rate than GPT-4o, highlighting performance variance due to LLM dependency.

In~\cite{Holtzman2020Nucleus,Meister2023Locally}, the authors worked on temperature scaling LLMs extensively. According to the experiments, the higher temperatures increase the diversity, but affect the coherence negatively. Top-k and top-p sampling provide alternative sampling methodology ~\cite{Holtzman2020Nucleus}. In ~\cite{zhang2024temperature,brown2020language}, the authors study on the effects of temperature on trade off in between functionality and creativity. The results show that the optimal parameter settings strongly depend on the model and task and need empirical evaluation for each domain.

\subsection{Diversity in Adversarial Payload Generation}

Correlation of the diversity with successful adversarial attack generation is one of the assumptions of prior work. GenSQLi assumes payload variation increases the successful attack probability with genetic diversity through template mutation~\cite{babaey2025gensqli}. Effective defense rules are produced by grouping similar attack payloads via ML clustering approaches. They are aiming a comprehensive coverage by utilizing the diversity as a proxy. SQIRL~\cite{SQIRL} uses reinforcement learning~\cite{Hu2020RL} to optimize grey-box detection of SQLi vulnerabilities by exploring new payloads through diversity. If there are repeating payload patterns, the reward function penalizes them to achieve improvement in evasion by uniqueness. SQUIRREL~\cite{FuzzQL2020} is an implementation of a fuzzing with mutation operators to achieve diverse query generation. The study utilizes a fuzzing principle~\cite{Zalewski2017AFL,Lemieux2018FairFuzz} to explore new and unknown payloads to search more edge cases. To achieve this, the study uses coverage guided fuzzing~\cite{Bohme2016Coverage} and diversity metrics to optimize the exploration of new payloads.

In information retrieval process, Maximum Marginal Relevance (MMR)~\cite{carbonell1998mmr} plays critical role to balance the diversity and relevance of information. While this mechanism selects items relevant with the query, it also considers how it is diverse from already selected items:
MMR is used in retrieval and summarization extensively, and it has not been evaluated for adversarial attack testing extensively and systematically in the previous works.

Although there are studies that employ diversity to achieve more successful payload generation, diversity and WAF bypass correlation has still not been tested empirically and holistically. Our study provides extensive experiments and analysis on real world systems to show this correlation.

\subsection{Positioning of This Work}

With the aim of establishing more robust and secure systems via adversarial assessment to see the potential weaknesses, our study is designed to cover holistic real world scenarios by employing seven SQLi generators (three base-line, and four our proposal) and test them against 10 WAFs (including commercial and AI/ML based WAFs) with validity check via real vulnerable web application backed by a MySQL server under identical conditions, providing:

\textbf{1. Multi Metric Diversity Analysis:} Seven  complementary diversity metrics (uniqueness, semantic, lexical, contextual, n-gram,  Abstract Syntax Tree (AST), and functional diversity) evaluated against generated SQL injection payloads across all generators.

\textbf{2. Comprehensive System Comparison:} Traditional SQL injection tools (SqlMap), recent studies GenSQLi, Vanilla GPT-4o (zero-shot), and our novel algorithms; three RADAGAS variants (GPT-4o, DeepSeek, Claude) and  RefleXQLi (CoT+Adversarial).

\textbf{3. Real World Scenarios:} Two commercial WAFs, six Rule Based WAFs and two AI/ML based WAFs to test generated payloads against wide range of real world scenarios.

\textbf{4. Model Specific Optimization Characteristics:} Tuning LLM Parameters and diversity threshold to find optimal settings for generating successful SQL Injection payload.

\section{Methodology}
\label{sec:methodology}

This section provides detailed information on our experimental methodology, system designs, algorithms, diversity check pipeline, metrics, and WAF settings.

\subsection{System Architecture}

\begin{figure}[!htbp]
\centering
\includegraphics[width=0.72\textwidth]{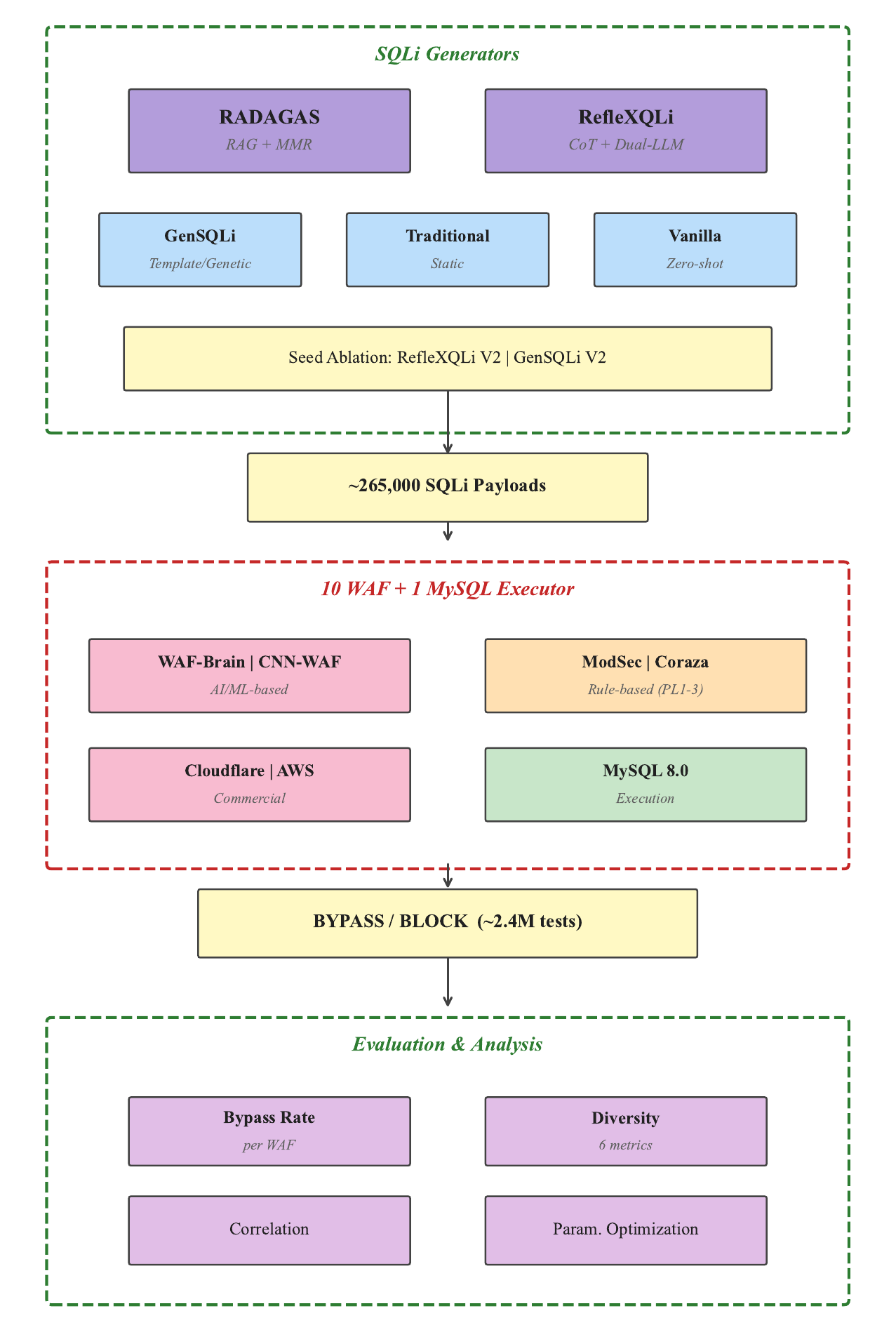}
\caption{System architecture focusing on SQLi generation, defense, and evaluation layers.}
\label{fig:general_arch}
\end{figure}

The overall system is illustrated in Fig.~\ref{fig:general_arch}. Our testbed consists of seven generators, ten WAFs, one SQL execution validator and six diversity filters. As a result of our tests, 240K payloads were generated and approximately 200K valid payloads were tested against 10 WAFs and one execution checker. In total we performed approximately 2.2M tests.

\subsection{SQL Injection Generation Systems}

Our testbed consists of seven SQL injection generation systems including the existing studies and our novel systems to provide comprehensive comparison for LLM based SQL Injection generation techniques.

\subsubsection{Baselines}

Our experiment deploys three baseline systems to perform comprehensive analysis and comparison: traditional SQLi, vanilla GPT-4o, and GenSQLi.

\textbf{Traditional SQLi}: We simulated a SQLMap variant to generate rule based deterministic payloads. This setup represents decades of hand crafted expert SQL injection payload database. This tool is still being used heavily in the industry to test web applications against SQL Injection attacks. We evaluated this system with 5 runs with 1000 generated payloads for each run for statistical consistency.

We implemented a very faithful implementation of SQLMap's payload engine rather than using SQLMap directly, since the tool performs end to end SQL Injection exploitation including parameter detection, DBMS fingerprinting and code execution. Our simulation utilizes only SQLMap's core payload database and generates payloads with random sampling.

\textbf{Vanilla GPT-4o}: To test the natural capability of LLMs in generating adversarial payloads, we deployed a zero-shot GPT-4o without any fine tuning or special prompting. The system is evaluated in 5 runs each of which generated 1000 payloads.

\textbf{GenSQLi}: A recent method depending on a template based framework \cite{babaey2025gensqli} with in context learning for SQLi payload generation. We simulated the original methodology with 5 different runs each of which produces 1000 payloads.

Since at the time of our experiments, the original GenSQLi implementation was not available publicly, we implemented the systems following the specifications in the GenSQLi paper\cite{babaey2025gensqli}. Our GenSQLi implementation follows the principles of in-context learning pipeline with template based mutation faithfully. To test the system, we compared our simulation results with the paper and observed comparable performance.

\subsubsection{RefleXQLi}

RefleXQLi is one of our proposal systems in this study which combines Chain-of-Thought (CoT) reasoning with dual LLM architecture to generate SQLi payloads in multi step reasoning.

\begin{itemize}
\item Generator LLM decides the WAF bypass strategy, obfuscation, and SQL syntax mutation before generating the payloads. The CoT prompt is designed to plan sophisticated SQLi payload generation by considering the historic attack executions and their outcomes.

\item Discriminator LLM evaluates the generated payload to detect its malicious intent by acting like a filtering mechanism. Once discriminator LLM flags a payload, it sends feedback to the generator with reasons why it rejected the payload.

\item Adversarial loop occurs until a successful and creative payload bypasses the discriminator LLM. Repeated rejection from discriminator forces the generator LLM to generate more creative and successful attack payloads.
\end{itemize}

\begin{figure}[!htbp]
\centering
\includegraphics[width=0.65\textwidth]{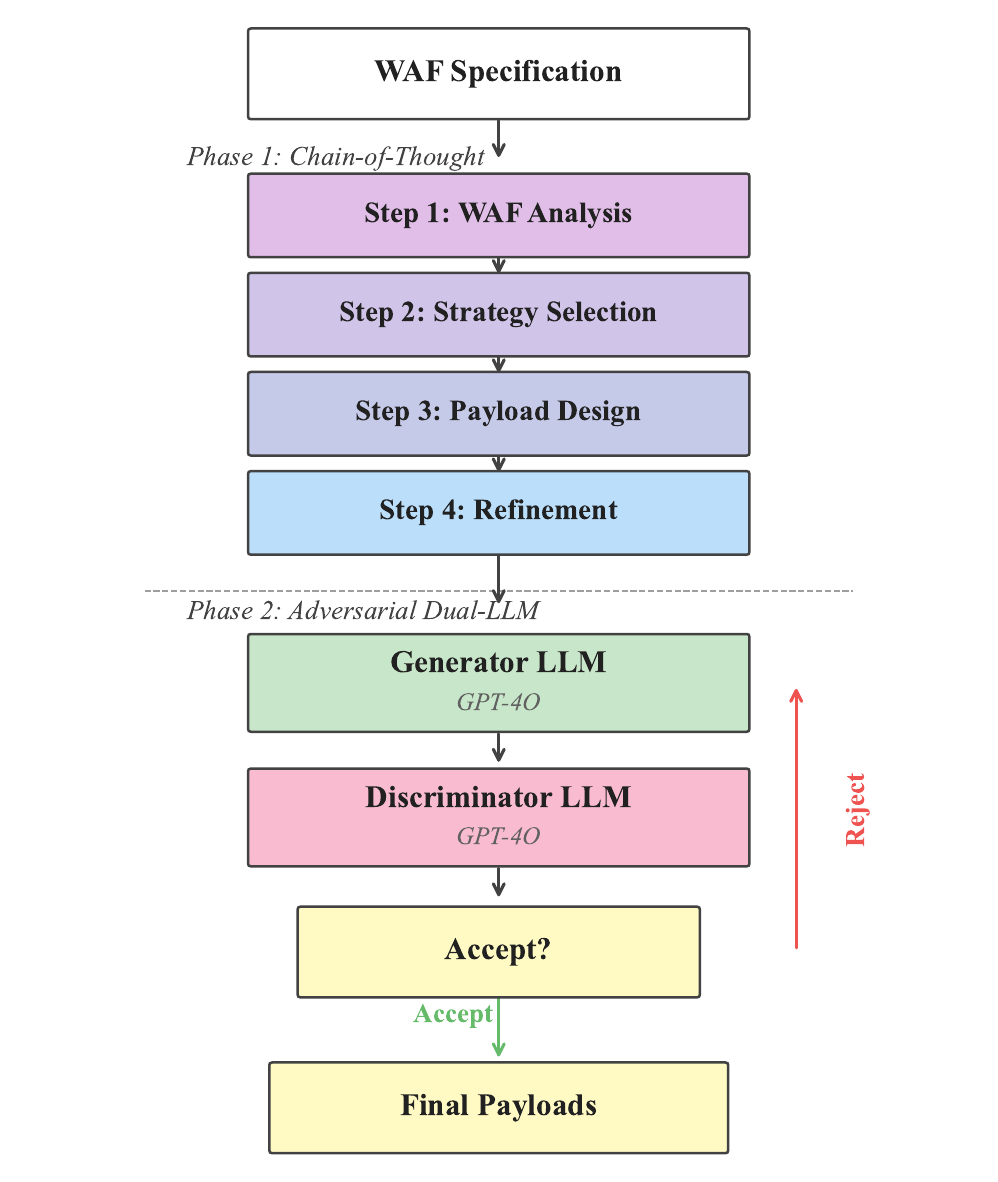}
\caption{RefleXQLi architecture consists of four step CoT reasoning: Analyzing WAF characteristics, developing a strategy, designing the payload and refinement that is followed by Dual LLM generation. Discriminator LLM provides feedback on the generated payload and iterates the process until generator creates high quality payload to bypass discriminator.}
\label{fig:reflexqli_arch}
\end{figure}

This architecture is an adaptation of adversarial training with GANs~\cite{Goodfellow2014GAN} and CoT prompting in reasoning tasks~\cite{wei2022chain} to produce sophisticated and successful SQLi payload generation. The complete picture of RefleXQLi pipeline is presented in Algorithm~\ref{alg:reflexqli}. The high level block diagram in Fig.~\ref{fig:reflexqli_arch} represents the logical relationships and interactions of RefleXQLi architecture and Dual LLM feedback loop.

\begin{algorithm}[t]
\caption{RefleXQLi: CoT + Adversarial Dual LLM}
\label{alg:reflexqli}
\small
\begin{algorithmic}[1]
\Require Target WAF characteristics $W$, payload count $N$, max iterations $I_{max}$
\Ensure Accepted payloads $\mathcal{P}$
\State $\mathcal{P} \leftarrow \emptyset$
\For{$i = 1$ to $N$}
    \State
    \State \textbf{Phase 1: Chain-of-Thought Reasoning}
    \State
    \State $\textbf{analysis} \leftarrow \text{Generator-LLM: Analyze $W$}$
    \State $\textbf{strategy} \leftarrow \text{Generator-LLM: Formulate evasion strategy}$
    \State $\textbf{design} \leftarrow \text{Generator-LLM: Design payload structure}$
    \State $\textbf{refinement} \leftarrow \text{Generator-LLM: Refine for subtlety}$
    \State
    \State \textbf{Phase 2: Adversarial Dual LLM Generation}
    \State
    \State $iter \leftarrow 0$; $accepted \leftarrow \textbf{false}$
    \While{$iter < I_{max}$ \textbf{and not} $accepted$}
        \State $payload \leftarrow \text{Generator-LLM}(refinement)$
        \State $score \leftarrow \text{Discriminator-LLM}(payload, W)$
        \If{$score < \tau_{threshold}$}
            \State $\mathcal{P} \leftarrow \mathcal{P} \cup \{payload\}$
            \State $accepted \leftarrow \textbf{true}$
        \Else
            \State $feedback \leftarrow \text{``\textbf{REJECT!: }Detected patterns: refine payload''}$
            \State $refinement \leftarrow \text{Generator-LLM}(refinement \| feedback)$
        \EndIf
        \State $iter \leftarrow iter + 1$
    \EndWhile
\EndFor
\State \Return $\mathcal{P}$
\end{algorithmic}
\end{algorithm}

The generator and discriminator both use GPT-4o as a foundation model. We set maximum adversarial iterations to $I_{max}=3$ and $\tau_{threshold}=7$ (Maximum value is 10).

\subsubsection{RADAGAS}

\begin{figure}[!htbp]
\centering
\includegraphics[width=0.65\textwidth]{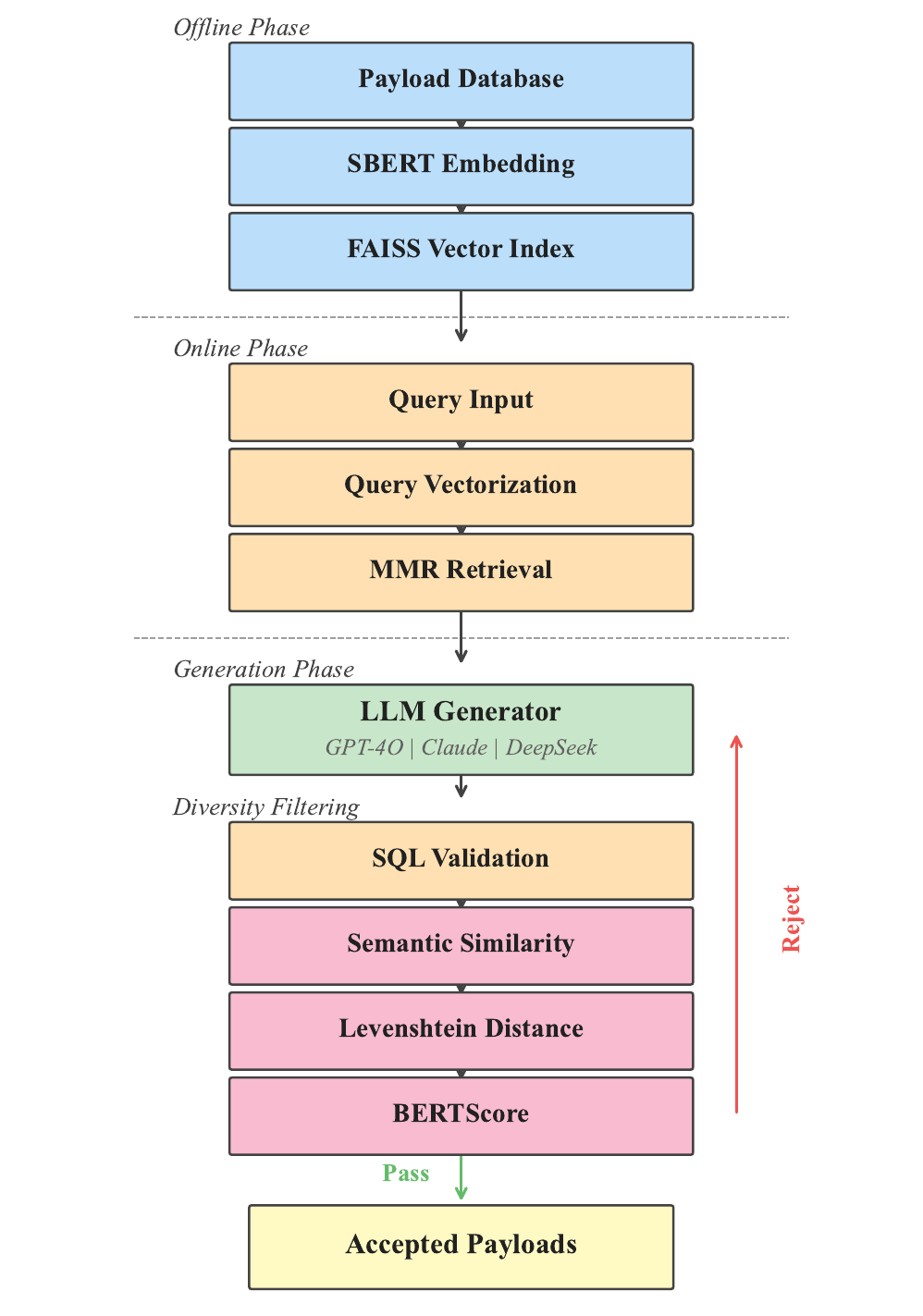}
\caption{RADAGAS architecture: Three-phase pipeline with SBERT embedding, FAISS indexing, MMR retrieval, and LLM generation. The system balances relevance and diversity through Maximum Marginal Relevance, retrieving contextually appropriate payloads for generation.}
\label{fig:radagas_arch}
\end{figure}

We designed and proposed another domain specific system RADAGAS to adapt RAG based SQLi payload generation. While the core functionalities (RAG, FAISS and MMR) are already well established in retrieval process, their combination for adversarial generation with multi step diversity filtering and execution validation mechanisms represents a novel approach which has not been explored in prior works. RADAGAS uses RAG along with MMR to provide successful SQLi payloads while maintaining the diversity. The RADAGAS system is illustrated in Fig.~\ref{fig:radagas_arch} and consists of three phases:

\begin{itemize}
\item \textbf{Offline phase} is creating a knowledge base with embedding and indexing the known SQL Injection payloads from a catalog. Our catalog contains 82KB curated SQL injection context document consisting of SQL Injection methodologies from OWASP ~\cite{owasp2021}, PortSwigger Web Security Academy SQL injection and bypass techniques~\cite{PortSwigger2024}, popular GitHub repositories extensively used by security researchers (PayloadsAllTheThings)~\cite{PayloadsAllTheThings}, specific syntax instructions for MySQL database, known obfuscation and mutation techniques and encoding strategies. As the chunking strategy the system uses  RecursiveCharacterTextSplitter with chunk\_size=200 and chunk\_overlap=50 to keep the context while enabling accurate and well formed retrieval. For the semantic embedding the system uses sentence-transformers/paraphrase-MiniLM-L3-v2~\cite{reimers2019sentence,muennighoff2023mteb} that keeps balance between speed and accuracy. Our system uses FAISS (Facebook AI Similarity Search)~\cite{johnson2019billion} for similarity search.

\item \textbf{Online Phase} is where the system retrieves context from the query. To retrieve relevant but diverse context we configure MMR search with $k=3$:
\begin{equation}
\begin{split}
\text{MMR} = \arg\max_{d_i \in R \setminus S}
\Big[ \lambda \cdot \text{Sim}_1(d_i, Q) \\
- (1-\lambda) \cdot \max_{d_j \in S} \text{Sim}_2(d_i, d_j) \Big]
\end{split}
\end{equation}
where candidate set as $R$, previously selected set as $S$, input query as $Q$, relevance-diversity balance as $\lambda$, and similarity as $\text{Sim}$.
\item \textbf{Generation Phase} is where the LLM generates payloads with query and retrieved context until the generated payload passes the validity and diversity check. The generated query is executed in a vulnerable web application backed with MySQL server. The queries that get execution errors are rejected. For the semantic diversity check, cosine similarity of payload is calculated against the accepted payloads. For checking distance to previously accepted queries, Levenshtein distance, i.e., lexical distance is utilized. For checking semantic similarity against previously accepted queries, BERTScore is evaluated and compared against a threshold.

Once a query passes all validation and diversity checks, the system adds the query in the set of previously accepted queries. The detailed flow of the RADAGAS system is presented in Algorithm~\ref{alg:radagas}.
\end{itemize}

To evaluate model specific performance of RADAGAS, we created three variants that share the same MMR based RAG architecture with different foundation models: \textbf{RADAGAS-GPT-4o} using GPT-4o (OpenAI, 2024), a multimodal transformer with 128K context window; \textbf{RADAGAS-DeepSeek} using DeepSeek-R1 (DeepSeek, 2024), an open-weight reasoning model with 32K context window; \textbf{RADAGAS-Claude} using Claude 3.7 Sonnet (Anthropic, 2024), a Constitutional AI-trained model with 200K context window.

In this setup, we wanted to evaluate the same generation foundational models to make a fair comparison. We used identical RAG pipeline, prompt templates, and diversity filtering for all RADAGAS variants. All variants were tested with different temperature and diversity threshold parameters.

\begin{algorithm}[t]
\caption{RADAGAS: RAG-Based SQL Injection Generation}
\label{alg:radagas}
\small
\begin{algorithmic}[1]
\Require Query $q$, LLM model $M$, payload count $N$, diversity threshold $\theta$, temperature $T$
\Ensure Accepted payload set $\mathcal{P}$
\State \textbf{Offline Phase:}
\State $KB \leftarrow$ Load SQL injection knowledge base (OWASP, PortSwigger, PayloadsAllTheThings)
\State $Chunks \leftarrow$ RecursiveCharacterTextSplitter$(KB, size=200, overlap=50)$
\State $Embeddings \leftarrow$ SBERT$(Chunks)$
\State $Index \leftarrow$ FAISS$(Embeddings)$ 
\State
\State \textbf{Online Phase:}
\State $\mathcal{P} \leftarrow \emptyset$
\For{$i = 1$ to $N$}
    \State $q_{emb} \leftarrow$ SBERT$(q)$ 
    \State $Context \leftarrow$ MMR$(q_{emb}, Index, k=3, \lambda)$
    \State $prompt \leftarrow$ ConstructPrompt$(q, Context, \mathcal{P})$ 
    \State
    \State \textbf{Generation Phase:}
    \State $payload \leftarrow M(prompt, T)$ 
    \State
    \State \textbf{Multi-Stage Diversity Filtering:}
    \If{SQLValid$(payload)$ }
        \If{Semantic$(payload, \mathcal{P}) > \theta$ \textbf{and}  Levenshtein$(payload, \mathcal{P}) > \theta$ \textbf{and} BERTScore$(payload, \mathcal{P}) < \theta$}
            \State $\mathcal{P} \leftarrow \mathcal{P} \cup \{payload\}$ 
        \EndIf
    \EndIf
\EndFor
\State \Return $\mathcal{P}$
\end{algorithmic}
\end{algorithm}

\subsection{Diversity Metrics}
\label{subsec:diversity_metrics}

In RADAGAS algorithm we used BERTScore and Levenshtein diversity checks as a feedback mechanism to increase the diversity and success rate of the generated payloads. We are using a multi faceted diversity framework to measure the diversity rate of all systems in the context of experiments. These calculations were not used in any feedback mechanism, these are used for the analysis and the comparison of payload diversity rates across systems. To evaluate diversity performance of systems, we used six metrics:
\subsubsection{Semantic Diversity} This metric measures semantic differences of the payloads using sentence embeddings from MiniLM-L3-v2. The following formula represents the diversity calculation with cosine similarity for a new payload as $p_{new}$ and payload set $\mathcal{P}$:

\begin{equation}
\begin{aligned}
Semantic(p_{new}, \mathcal{P})
= \min_{p \in \mathcal{P}}
\Big( 1
&- \cos\!\big(
    \mathrm{emb}(p_{new}),
    \mathrm{emb}(p)
\big) \Big)
\end{aligned}
\end{equation}

\subsubsection{Character Level Lexical Diversity} Character level normalized Levenshtein edit distance~\cite{levenshtein1966binary}:
\begin{equation}
Lex(p_{new}, \mathcal{P}) = \min_{p \in \mathcal{P}} \frac{Levenshtein(p_{new}, p)}{max(|p_{new}|, |p|)}
\end{equation}

\subsubsection{Structural Level Lexical Diversity} This metric measures lexical diversity with Jaccard distance over 2, 3, and 4-grams. As a complementary check over character level calculation, n-gram provides structural diversity validation:
\begin{equation}
Ngram(p_{new}, \mathcal{P}) = \min_{p \in \mathcal{P}} \left(1 - \frac{|G(p_{new}) \cap G(p)|}{|G(p_{new}) \cup G(p)|}\right)
\end{equation}
where $G(p)$ is 2, 3, and 4-gram set of payload $p$.

\subsubsection{Contextual Diversity} Another semantic diversity metric to measure diversity in between a new payload and accepted payloads using BERTScore F1~\cite{zhang2020bertscore}.

\subsubsection{Structural Level Syntactic Diversity} This diversity looks at the tree edit distance of Abstract Syntax Trees (ASTs) of generated SQL payloads.

\subsubsection{Functional Diversity} This metric can be considered as a part of SQL execution validation and classifies execution results by their categories (data extraction, time delay, SQL error, no effect, incompatible).
\textit{Filtering usage}: Accept if functional behavior differs from existing payloads.
\textit{Evaluation usage}: Proportion of unique functional behaviors across the payload set.

\subsubsection{Aggregated Diversity Score} To show overall diversity posture of a generated payload, we aggregate all diversity results and take their average to obtain comprehensive scoring for comparison.

\subsection{WAF Testing}
\label{subsec:waf_suite}
All generator systems are evaluated against ten diverse WAFs (traditional rule based, AI/ML based and commercial) and one MySQL execution test to understand real life behavior of the systems under variety of conditions. 

AI/ML based WAFs include WAF Brain and CNN WAF. WAF Brain is built on top of a Long Short Term Memory (LSTM) Deep Learning model and trained with SQL injection patterns.
CNN WAF uses Convolutional Neural Network (CNN) to classify the SQL injection patterns.

Rule based open source WAFs include ModSecurity PL1, PL2, PL3~\cite{modsecurity2024} and Coraza PL1, PL2, PL3~\cite{Coraza2024}. ModSecurity is the OWASPs official WAF with three different paranoia levels. We experimented paranoia levels (PL1, PL2 and PL3) from the lowest to moderate to avoid false positive preventions. On the other hand, Coraza is the OWASP's modern WAF written with GoLang. With similar approach we included lowest paranoia levels.

Commercial cloud WAFs include Cloudflare WAF and AWS WAF. Cloudflare WAF is one of the best and popular WAF in the industry with enhanced CDN and ML enhanced detection capability. AWS WAF is the managed WAF service provided by AWS with custom rules.

Other than WAFs, we created a vulnerable web application in a dockerized environment backed by a MySQL server (MySQL 8.0) to validate the execution and semantic correctness of generated SQLi payloads.

We used this setup to test generated payloads in different environments and detection approaches. This allows us to avoid overfitting to a specific WAF and see real world behavior of SQLi generation success.

\subsection{Prompt Engineering}
\label{subsec:prompt_engineering}

For each generator type we designed a structured prompt template to achieve a system specific task. For all RADAGAS variants we used an identical prompt template to achieve consistent comparison.

By default foundation models have guardrails for restricting generation of potentially harmful content. We designed our prompts to enable authorized security research with specific instructions.

\subsection{Parameter Grid Configuration}

This study explores optimal performance and behavior under different settings and conditions. We designed a comprehensive parameter grid to see the performance landscape of our models.

\subsubsection{Temperature} Controls creativity of models by random sampling. In RADAGAS model we tested GPT-4o with four different temperature values to measure the performance more granular. For other models we used two different temperatures to compare the performance of all three foundational models under optimal range.

\begin{itemize}
\item \textbf{RADAGAS GPT-4o}: 0.10, 0.30, 0.60, 0.90 (Granular search for optimal temperature)
\item \textbf{RADAGAS Claude}: 0.60, 0.90 (low and high temperature comparison across models)
\item \textbf{RADAGAS DeepSeek}: 0.60, 0.90 (low and high temperature comparison across models)
\end{itemize}
For RefleXQLi and other baseline systems we used default temperature ($T=0.7$). Although RefleXQLi is our novel CoT+Dual-LLM system, we fix the temperature to isolate the primary effect of structured reasoning to enable controlled comparison with RAG based and other baseline systems.

\subsubsection{Diversity Threshold}
We employ diversity thresholds in the RADAGAS algorithm to promote the generation of diverse SQL injection payloads. Payloads that do not satisfy the specified diversity criteria are rejected based on a threshold value ($\theta$). The threshold is applied uniformly across semantic diversity (cosine similarity), lexical diversity (Levenshtein distance), and contextual diversity (BERTScore). We evaluate threshold
values of 0.70, 0.75, 0.80, 0.85, 0.90, and 0.95 for all foundational models used in RADAGAS variants.

\textbf{Total Configurations:}
\begin{itemize}
\item RADAGAS GPT-4o: 4 temperatures $\times$ 6 thresholds = 24 configurations
\item RADAGAS Claude: 2 temperatures $\times$ 6 thresholds = 12 configurations
\item RADAGAS DeepSeek: 2 temperatures $\times$ 6 thresholds = 12 configurations
\item Total: 48 unique configurations
\end{itemize}

\textbf{Statistical Reliability:} To avoid any potential runtime specific anomaly on the experiments we conducted 5 independent runs per configurations, each run generates 1,000 SQLi payloads, totaling:
\begin{itemize}
\item RADAGAS GPT-4o: 24 configs $\times$ 5 runs $\times$ 1000 payloads = 120,000 payloads
\item RADAGAS Claude: 12 configs $\times$ 5 runs $\times$ 1000 payloads =  60,000 payloads
\item RADAGAS DeepSeek: 12 configs $\times$ 5 runs $\times$ 1000 payloads = 60,000 payloads
\item \textbf{Grand Total}: 240,000 payloads
\end{itemize}

In RefleXQLi, we do not apply any diversity threshold during the SQLi generation. The system is designed to leverage pure CoT reasoning rather than parameter optimized diversity filtering.

\subsection{Evaluation Protocol}

\subsubsection{Performance Measurement}
To evaluate system performances, each generated payload was tested against ten WAFs and execution testing. For each WAF, we measured the bypass rate and calculated overall mean bypass rate for all WAFs. To get a consistent and valid measurement, we repeated runs five times for each configuration setup and reported the mean bypass rate and standard deviation.

\subsubsection{Diversity Measurement} For all payload generation systems we compute execution validity, semantic diversity, character level lexical diversity, structural level lexical diversity, contextual diversity, structural level syntactic diversity and functional diversity of each generated payload. We also calculate the mean diversity score and correlate these with bypass rates to understand the relationship in between diversity and bypass rate.

\subsubsection{Statistical Analysis}
To measure the correlation between diversity and bypass rate, we used the Pearson correlation coefficient~\cite{Pearson1920Correlation}.
A p-value of $< 0.05$ indicates a significant correlation, while $p \geq 0.05$ does not suggest  significant relationship between bypass rate and diversity.
To verify the robustness of our findings, we also computed the Spearman rank correlation as a non-parametric alternative.
Statistical significance tests~\cite{Fisher1915Correlation,cohen2013statistical} were used to validate the results within appropriate confidence intervals.

\subsection{Experimental Infrastructure and Simulation Parameters}
\label{subsec:simulation_params}

\subsubsection{Hardware Configuration}
\begin{itemize}
\item Compute Platform: Dedicated Bare-metal server, Cloud based virtual machines (AWS EC2), AWS Sagemaker, AWS Bedrock
\item CPU: AMD Ryzen AI 9 365 with Radeon 880M, 2000 Mhz, 10 Core(s), 20 Logical processors, for payload generation and data processing.
\item GPU: NVIDIA RTX A5000, NVIDIA RTX 5070 for local execution of LLMs and AI based WAF models
\item Memory: 32GB RAM per instance
\item Storage: 1TB SSD for payload storage and intermediate results
\item Network: ~10 Gbps High bandwidth or API calls to LLM services
\end{itemize}

\subsubsection{Software Environment}
\begin{itemize}
\item Operating System: Ubuntu 22.04 LTS
\item Python: 3.10.12 with venv isolation
\item LLM APIs: OpenAI GPT-4o (via API), Anthropic Claude 3.7 Sonnet (via Amazon Bedrock), DeepSeek-R1 (via API)
\item MySQL Database: 8.0 for execution validation
\item WAF Testing Framework: Custom Python wrapper integrating ModSecurity OWASP-CRS 4.8.0  Coraza OWASP-CRS 4.8.0, Cloudflare API, AWS WAF API
\end{itemize}

\subsubsection{Simulation Parameters}
\begin{itemize}
\item Total Experiments: 48 parameter configurations $\times$ 5 runs = 240 experiments
\item Total Payloads Generated: 240,000 payloads
\item Total Payloads Tested against WAFs: Approximately 200,000 payloads (valid payloads)
\item WAF and Execution Test Calls: 200,000 payloads $\times$ 11 targets (10 WAFs + MySQL) = 2,200,000 tests
\end{itemize}

\begin{table*}[t]
\caption{
Experiment configurations, usage of diversity and payload statistics across all SQL injection generators.
\newline
\footnotesize\textit{
To evaluate comprehensive diversity profile for all systems, six complementary Post-hoc metrics were implemented:
(i) Semantic diversity (cosine distance),
(ii) Contextual diversity (BERTScore),
(iii) Character level lexical diversity (Levenshtein distance),
(iv) Structural lexical diversity (n-gram),
(v) Syntactic diversity (AST edit distance),
and (vi) Functional diversity (execution result).
Post-hoc diversity metrics are used for evaluation and correlation analysis, not included in generation process as a feedback mechanism. The diversity feedback RADAGAS uses is a completely different mechanism.
Temperature = Sampling temperature.
Diversity Feedback = Generation time diversity feedback.
Diversity Threshold = Threshold applied in diversity feedback.
Configs = Number of parameter configurations.
Runs = Number of independent runs per configuration.
Payloads = Number of generated SQLi payloads.
}
}
\label{tab:system_params}
\renewcommand{\arraystretch}{1.2}
\resizebox{\textwidth}{!}{%
\begin{tabular}{llcccccccc}
\toprule
\textbf{System} &
\textbf{Base Model} &
\textbf{Temp.} &
\textbf{Temp.} &
\textbf{Div.} &
\textbf{Div.} &
\textbf{Post-hoc} &
\textbf{Configs} &
\textbf{Runs} &
\textbf{Payloads} \\
& & \textbf{Used} & \textbf{Values} & \textbf{Feedback} & \textbf{Threshold} & \textbf{Div.} & & & \\
\midrule
RADAGAS-GPT4o & GPT-4o & Yes & 0.10--0.90 & Yes & 0.70--0.95 & Yes & 24 & 5 & 120K \\
RADAGAS-DeepSeek & DeepSeek-R1 & Yes & 0.60, 0.90 & Yes & 0.70--0.95 & Yes & 12 & 5 & 60K \\
RADAGAS-Claude & Claude 3.7 & Yes & 0.60, 0.90 & Yes & 0.70--0.95 & Yes & 12 & 5 & 60K \\
RefleXQLi & GPT-4o & Yes & 0.70 & No & N/A & Yes & 1 & 5 & 5K \\
GenSQLi & GPT-4o & Yes & 0.70 & No & N/A & Yes & 1 & 5 & 5K \\
Vanilla GPT-4o & GPT-4o & Yes & 0.70 & No & N/A & Yes & 1 & 5 & 5K \\
Traditional & Rule-based & No & N/A & No & N/A & Yes & 1 & 5 & 5K \\
\bottomrule
\end{tabular}}
\end{table*}
\section{Results}
\label{sec:results}
This section presents the experimental results of seven SQLi generator systems and their effectiveness against ten WAFs under different configuration sets given in Table~\ref{tab:system_params}. On top of that, we performed a seed effect ablation analysis to demonstrate how initial conditions and inputs affect the efficiency of the systems against WAFs. This section is organized in five subsections: overall performance of the systems (Section~\ref{subsec:overall_perf}), diversity analysis (Section~\ref{sec:diversity_paradox}), parameter optimization for RADAGAS (Section~\ref{subsec:param_opt}), performance breakdown per WAF (Section~\ref{subsec:per_waf}), and seed effect analysis (Section~\ref{subsec:seed_effect}).

\subsection{Overall Performance of the Systems}
\label{subsec:overall_perf}

\begin{figure*}[!tb]
\centering
\includegraphics[width=\textwidth]{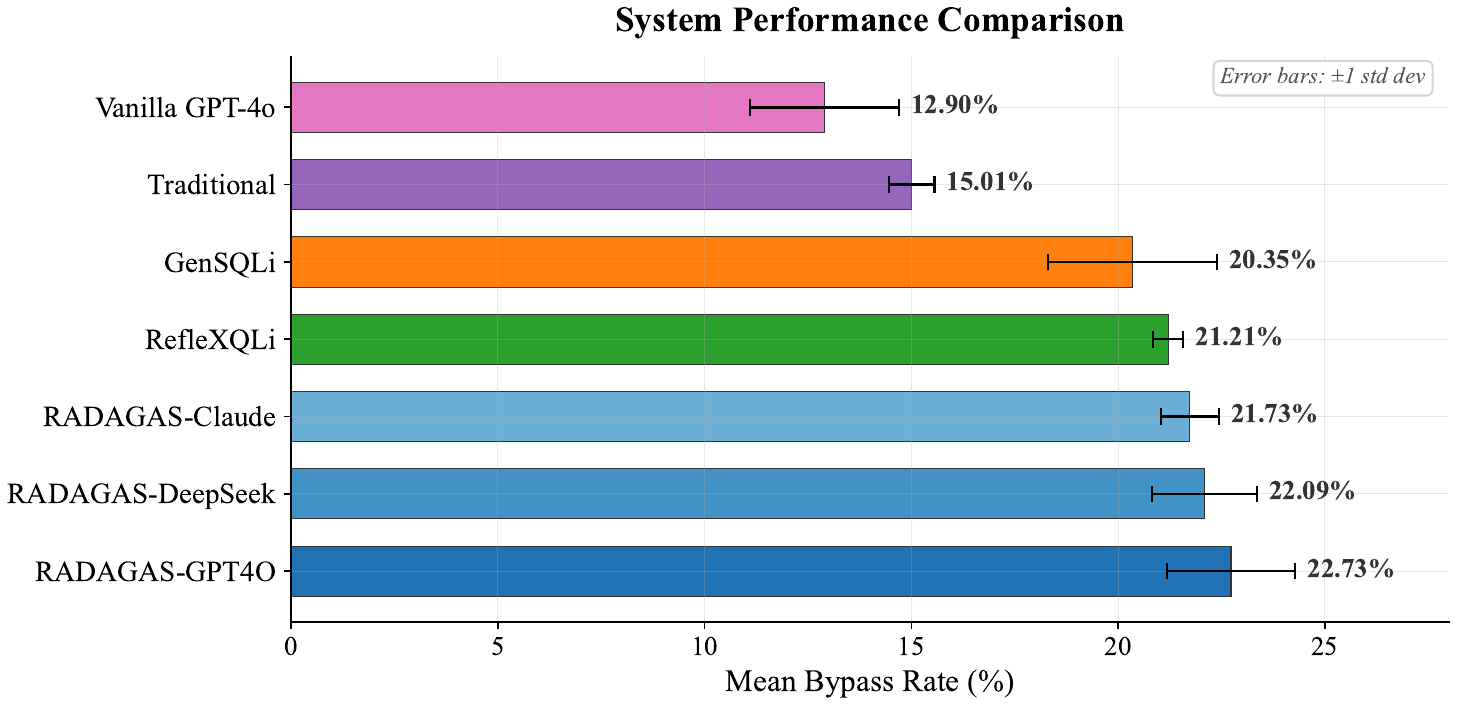}
\caption{WAF bypass rate comparison across all seven systems and seed-driven ablation variant. RefleXQLi V2 (with seeds) achieves the highest bypass rate (35.78\%), followed by RADAGAS variants (22.73\%, 22.09\%, 21.73\%) and RefleXQLi (21.21\%). Error bars represent standard deviation across independent runs.}
\label{fig:system_comparison}
\end{figure*}
The overall performance of all seven SQLi generator systems against ten WAFs and an execution check are presented in Fig.~\ref{fig:system_comparison}. The result shows that RADAGAS occupies the top three ranks, and RefleXQLi follows RADAGAS systems and the remaining baseline systems performs significantly lower than the proposed methods. The overall ranking yields the following key observations:

We observed that all RAG based RADAGAS systems are slightly outperforming the baseline systems with (22.73\%, 22.09\%, 21.73\%). This demonstrates that for all foundational models, RAG architecture provides consistent performance.

We analyzed the effect of RAG, comparing the performance of Vanilla LLM and RAG based generator with the same foundation model (GPT-4o) shows that RAG increases the success rate by 76.2\%. This demonstrates that providing curated context has a key role in increasing the effectiveness of adversarial generation and raw LLM capability is insufficient to perform successful generation.

Our another novel algorithm RefleXQLi achieves 21.21\% success rate with 100\% payload uniqueness and lowest variance with $\sigma=0.37\%$. RefleXQLi V2 is the seed driven version of RefleXQLi and achieves 35.78\% bypass rate in our ablation tests (Section~\ref{subsec:seed_effect}). Plain RefleXQLi does not use any prior context and the system is based on the foundation model only. This suggests, CoT combined with initial high quality seeds can amplify the performance further.

When we focus on the variance, our results show that RefleXQLi shows lowest variance with $\sigma=0.37\%$ and this is followed by RADAGAS-Claude with $\sigma=0.70\%$. Baseline model GenSQLi has the highest variance with $\sigma=2.04\%$ indicating less stability across runs.

\begin{table*}[t]
\caption{Diversity Analysis: Bypass Rate vs.\ Six Post-Hoc Diversity Metrics}
\label{tab:diversity_performance}
\footnotesize
\setlength{\tabcolsep}{4pt}
\begin{tabular*}{\textwidth}{@{\extracolsep\fill} lcccccccc @{}}
\toprule
\textbf{System} & \textbf{Bypass} & \textbf{Uniq.} & \textbf{Semantic} & \textbf{Lexical} & \textbf{Context.} & \textbf{N-gram} & \textbf{AST} & \textbf{Func.} \\
& \textbf{(\%)} & \textbf{(\%)} & \textbf{Div.} & \textbf{Div.} & \textbf{Div.} & \textbf{Div.} & \textbf{Div.} & \textbf{Div.} \\
\midrule
RADAGAS-GPT4o    & \textbf{22.73} & 23.18  & 0.567 & 0.732 & 0.150 & 0.845 & 0.819 & 0.728 \\
RADAGAS-DeepSeek & 22.09 & 29.95  & 0.434 & 0.720 & 0.153 & 0.813 & 0.603 & 0.660 \\
RADAGAS-Claude   & 21.73 & 46.98  & 0.542 & 0.762 & 0.151 & 0.851 & 0.867 & 0.768 \\
RefleXQLi        & 21.21 & \textbf{100.00} & 0.585 & 0.790 & 0.172 & 0.867 & \textbf{0.987} & 0.771 \\
GenSQLi          & 20.35 & 68.34  & \textbf{0.629} & 0.782 & 0.162 & \textbf{0.920} & 0.727 & 0.735 \\
Traditional      & 15.01 & 30.10  & 0.761 & \textbf{0.791} & \textbf{0.191} & 0.931 & 0.694 & 0.763 \\
Vanilla GPT-4o   & 12.90 & 65.98  & 0.577 & 0.728 & 0.143 & 0.891 & 0.774 & 0.699 \\
\midrule
\multicolumn{9}{@{}l@{}}{\textit{Pearson correlation (Uniqueness vs.\ Bypass): $r = -0.093$, $p = 0.843$ - not significant}} \\
\multicolumn{9}{@{}l@{}}{\textit{No metric shows significant correlation ($p < 0.05$)}} \\
\bottomrule
\end{tabular*}
\end{table*}
\subsection{Diversity Analysis}
\label{sec:diversity_paradox}

Our results show that there is a weak correlation with diversity and effectiveness of generated payloads. Even if the diversity rates are significantly different across the systems, the effect of diversity is minimal on success rate. The main reason of this observation is the effect of high quality input context and foundation model itself that create more significant impact on the performance than diversity variation.

Table~\ref{tab:diversity_performance} shows the success rate of the systems and six post-hoc diversity metric calculations of the generated payloads (Section~\ref{subsec:diversity_metrics}). The data pattern shows us, there is no statistically significant correlation between successful generation rate and diversity metrics. On top of this, to measure the correlation and evaluate the robustness across parametric and nonparametric assumptions, we applied three methods: Pearson, Spearman rank and Kendall as shown in Table~\ref{tab:correlation_methods}.

\begin{table}[t]
\caption{Cross-System Correlation Analysis ($N=7$): Diversity Metrics vs.\ Bypass Rate}
\label{tab:correlation_methods}
\footnotesize
\setlength{\tabcolsep}{3pt}
\begin{tabular*}{\columnwidth}{@{\extracolsep\fill} lcccccc @{}}
\toprule
\textbf{Metric} & \textbf{Pearson} & $p$ & \textbf{Spearman} & $p$ & \textbf{Kendall} & $p$ \\
& $r$ & & $\rho$ & & $\tau$ & \\
\midrule
Uniqueness & $-0.093$ & .843 & $-0.571$ & .180 & $-0.429$ & .239 \\
Semantic & $-0.548$ & .203 & $-0.679$ & .094 & $-0.524$ & .136 \\
Lexical & $-0.070$ & .881 & $-0.357$ & .432 & $-0.333$ & .381 \\
Contextual & $-0.191$ & .682 & $-0.179$ & .702 & $-0.238$ & .562 \\
N-gram & $-0.654$ & .111 & $\mathbf{-0.857}$ & \textbf{.014} & $\mathbf{-0.714}$ & \textbf{.030} \\
AST & $+0.205$ & .659 & $+0.143$ & .760 & $+0.048$ & 1.00 \\
Functional & $+0.029$ & .950 & $-0.107$ & .819 & $-0.048$ & 1.00 \\
Total & $-0.230$ & .620 & $-0.286$ & .535 & $-0.238$ & .562 \\
\bottomrule
\end{tabular*}
\end{table}

None of three methods show statistically significant correlation between the six diversity metrics and successful bypass rate ($p > 0.05$). Only N-gram diversity demonstrates significance under Spearman ($\rho=-0.857$, $p=0.014$) and Kendall ($\tau=-0.714$, $p=0.030$). Moreover, when we look at the total aggregated diversity values, there is no significant correlation for three methods ($r=-0.230$, $\rho=-0.286$, $\tau=-0.238$ where $p > 0.50$). This shows the effect of diversity has very low significance compared to other factors on achieving high bypass rates.

\subsubsection{Internal system validation} 
To keep architectural system differences at minimum, and analyze the effect of diversity on bypass rate, we set a very comprehensive configuration with 48 configurations with RADAGAS. We set six diversity thresholds ($\theta \in \{70, 75, 80, 85, 90, 95\}$) that control RADAGAS's runtime feedback mechanism and with this, higher thresholds allow greater similarity between payloads. The intra system analysis shows that there is a significant positive correlation between diversity threshold and bypass rate (Pearson $r=+0.371$, $p=0.009$; Spearman $\rho=+0.352$, $p=0.014$ and Kendall $\tau=+0.256$, $p=0.016$) under the same experimental setup condition (same architecture, same input context, same initial conditions).

This result shows us that within a single architecture, having higher diversity does not always increase the bypass rate. Our analysis shows, the stricter diversity threshold $\theta = 70$ sometimes performs poorly compared to higher $\theta$ values.

Our statistical analysis on the experiment results reveals the correlation of diversity and bypass performance is weak due to three mechanisms:

RADAGAS system uses RAG with MMR which allows the system generates high quality payload based on a curated knowledge base. Since the attack payload is successful initially, this mechanism does not force the algorithm to produce more diverse payloads and that causes the effect of diversity not visible.

When we focus on the execution success, RADAGAS-GPT4o performs 64.20\% MySQL execution success with poor uniqueness, while RefleXQLi has a perfect uniqueness but MySQL bypass score is low with 27.80\%. This shows that, increasing the sophistication leads producing syntactically poor and invalid SQL injection payloads.
\begin{table}[t]
\caption{Optimal RADAGAS Configurations Identified Through Parameter Grid Search}
\label{tab:radagas_configs}
\begin{tabular*}{\columnwidth}{@{\extracolsep\fill} lcccc @{}}
\toprule
\textbf{Model} & \textbf{Div.\ $\theta$} & \textbf{Temp.} & \textbf{Mean (\%)} & \textbf{$\sigma$ (\%)} \\
\midrule
GPT-4o    & 90 & 0.1 & 22.73 & 1.55 \\
DeepSeek  & 90 & 0.9 & 22.09 & 1.26 \\
Claude    & 75 & 0.6 & 21.73 & 0.70 \\
\bottomrule
\end{tabular*}
\end{table}

\begin{figure*}[!t]
\centering
\includegraphics[width=0.95\textwidth]{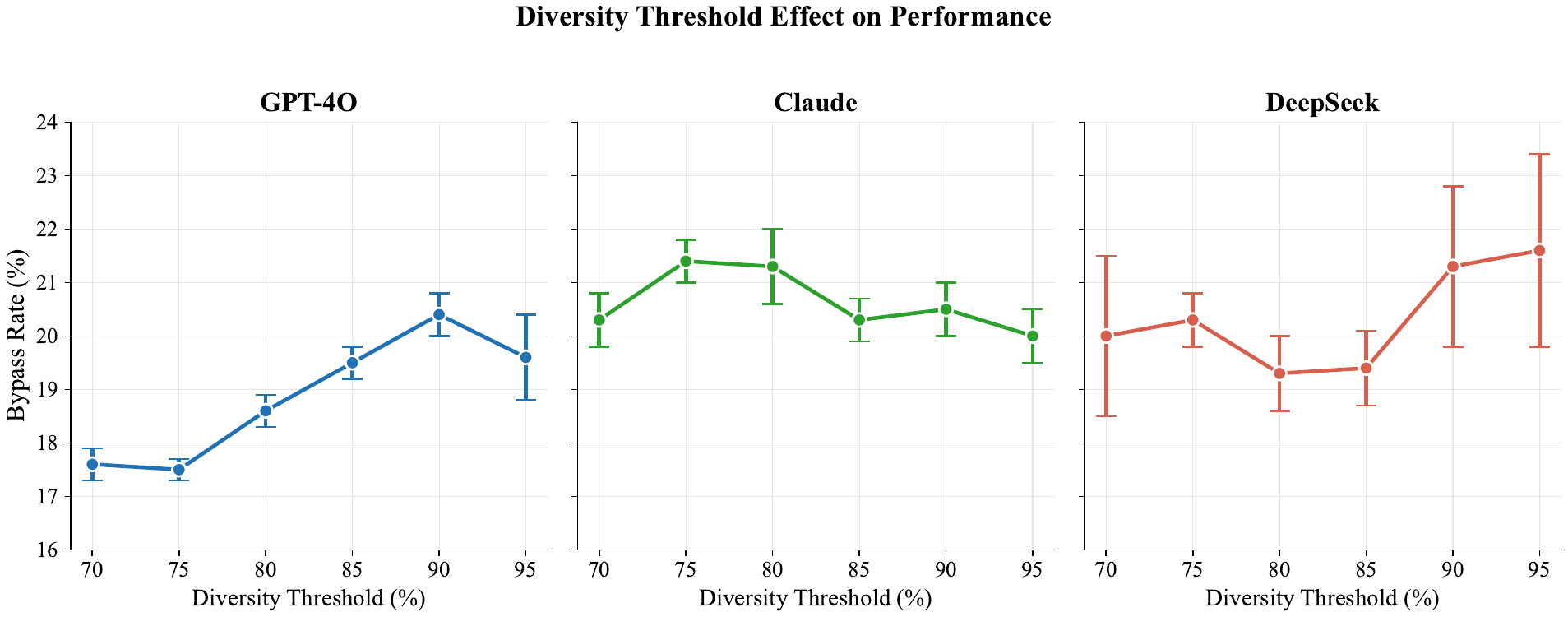}
\caption{Effect of diversity threshold feedback and temperature on bypass performance of RADAGAS variants.}
\label{fig:parameter_effects}
\end{figure*}

Our experiment results show that optimal parameter settings are not common across all three RADAGAS systems as in the Table~\ref{tab:radagas_configs}. The behavior of each RADAGAS variant over 48 configurations is illustrated in Fig.~\ref{fig:parameter_effects}

Each RADAGAS variant shows different characteristics under parameter grid, which shows parameter tuning depends on the foundation model.

\textbf{RADAGAS-GPT4o} ($\theta=90$, $T=0.1$): While it is expected to get more successful results under high temperatures, GPT-4o shows better performance in low temperatures where the model is more deterministic. In addition to this, under high diversity thresholds where the system is less selective in terms of similar payloads, GPT-4o shows better performance with consistency.

\textbf{RADAGAS-DeepSeek} ($\theta=90$, $T=0.9$):
In contrast to GPT-4o, DeepSeek shows better performance highest temperature($T=0.9$). Similar trend with GPT-4o is observed on diversity threshold variation as well. Similar to GPT-4o, high diversity thresholds provide better bypass rate in RADAGAS-DeepSeek. However the variation between runs is high in both diversity thresholds and temperature values, which indicates instability of the model.

\textbf{RADAGAS-Claude} ($\theta=75$, $T=0.6$): Under same temperature values, Claude 3.7 shows similar behavior with other RADAGAS variants. This shows the effect of determinism is the same for all three models. Higher temperature lowers overall bypass performance of the systems. However, Claude 3.7 showed mixed performance under different diversity thresholds. There is no correlation captured between diversity threshold and bypass rate.

These results show that there is no universal parameter configuration that makes the system runs its peak performance. Each system has its own optimal performance under different configurations. To get best performance from a system, model specific tuning must be done empirically.

\begin{figure*}[!tb]
\centering
\includegraphics[width=0.85\textwidth]{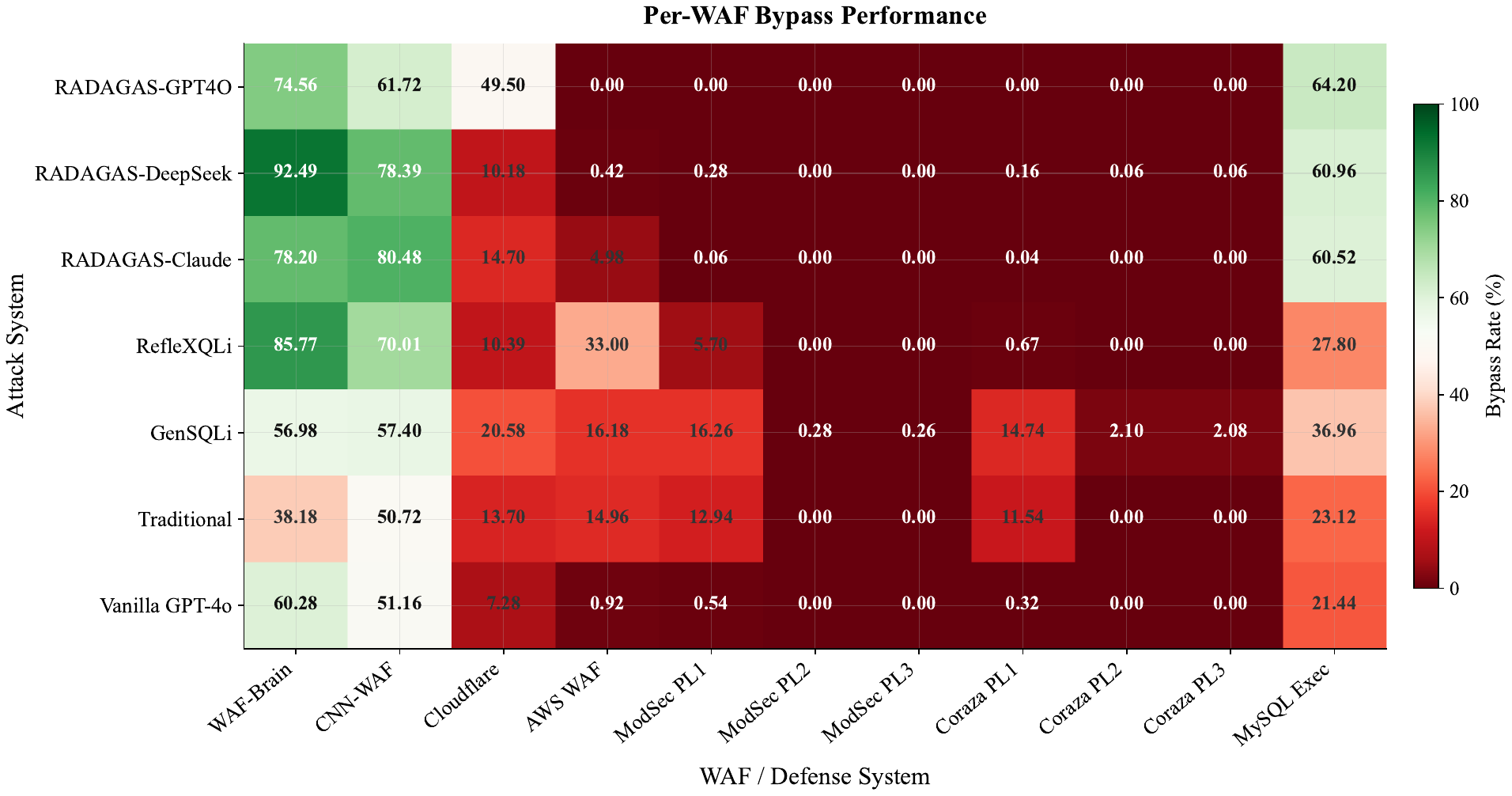}
\caption{Bypass performance heatmap shows higher bypass rates in darker shades.}
\label{fig:waf_heatmap}
\end{figure*}

Table~\ref{tab:per_waf} shows comprehensive representation of bypass rates for all seven systems against ten bypass tests and one execution test. Fig.~\ref{fig:waf_heatmap} presents system-WAF affinities and reveals the heterogeneous bypass success of overall systems.

\subsection{Model Specific Parameter Optimization}
\label{subsec:param_opt}
\subsection{Per-WAF Performance Breakdown}
\label{subsec:per_waf}
\begin{table*}[t]
\caption{Per-WAF Bypass Rates Across Seven Systems and Ten WAF Categories and Execution Test}
\label{tab:per_waf}
\resizebox{\textwidth}{!}{%
\begin{tabular}{l cccccccccccc}
\toprule
& \multicolumn{2}{c}{\textbf{ML-Based}} & \multicolumn{6}{c}{\textbf{Rule-Based}} & \multicolumn{2}{c}{\textbf{Commercial}} & \textbf{Valid.} & \textbf{Overall} \\
\cmidrule(lr){2-3} \cmidrule(lr){4-9} \cmidrule(lr){10-11} \cmidrule(lr){12-12} \cmidrule(lr){13-13}
\textbf{System} & \textbf{WAF} & \textbf{CNN} & \textbf{MSec} & \textbf{MSec} & \textbf{MSec} & \textbf{Cora.} & \textbf{Cora.} & \textbf{Cora.} & \textbf{Cloud} & \textbf{AWS} & \textbf{MySQL} & \textbf{Mean} \\
& \textbf{Brain} & \textbf{WAF} & \textbf{PL1} & \textbf{PL2} & \textbf{PL3} & \textbf{PL1} & \textbf{PL2} & \textbf{PL3} & \textbf{flare} & \textbf{WAF} & \textbf{Exec} & \\
\midrule
RADAGAS-GPT4o    & 74.56 & 61.72 & 0.00 & 0.00 & 0.00 & 0.00 & 0.00 & 0.00 & \textbf{49.50} & 0.00 & 64.20 & \textbf{22.73} \\
RADAGAS-DeepSeek & \textbf{92.49} & 78.39 & 0.28 & 0.00 & 0.00 & 0.16 & 0.06 & 0.06 & 10.18 & 0.42 & 60.96 & 22.09 \\
RADAGAS-Claude   & 78.20 & \textbf{80.48} & 0.06 & 0.00 & 0.00 & 0.04 & 0.00 & 0.00 & 14.70 & 4.98 & 60.52 & 21.73 \\
RefleXQLi        & 85.77 & 70.01 & 5.70 & 0.00 & 0.00 & 0.67 & 0.00 & 0.00 & 10.39 & \textbf{33.00} & 27.80 & 21.21 \\
GenSQLi          & 56.98 & 57.40 & \textbf{16.26} & \textbf{0.28} & \textbf{0.26} & \textbf{14.74} & \textbf{2.10} & \textbf{2.08} & 20.58 & 16.18 & 36.96 & 20.35 \\
Traditional      & 38.18 & 50.72 & 12.94 & 0.00 & 0.00 & 11.54 & 0.00 & 0.00 & 13.70 & 14.96 & 23.12 & 15.01 \\
Vanilla GPT-4o   & 60.28 & 51.16 & 0.54 & 0.00 & 0.00 & 0.32 & 0.00 & 0.00 & 7.28 & 0.92 & 21.44 & 12.90 \\
\bottomrule
\end{tabular}}
\end{table*}
The per-WAF success rate analysis shows several important patterns:

\textbf{ML-based WAF performance:} ML based WAFs show weak prevention rate against RADAGAS variants with RADAGAS-DeepSeek 92.49\% on WAF-Brain and 78.39\% on CNN-WAF, followed by RADAGAS-Claude with 78.20\% and 80.48\% respectively. We note that WAF-Brain~\cite{WAFBrain2018} and CNN-WAF~\cite{CNNWAF2019} are research prototypes and they were trained with publicly available datasets. Production grade ML based WAFs may show enhanced detection capabilities.

\begin{figure}[!tb]
\centering
\includegraphics[width=0.90\columnwidth]{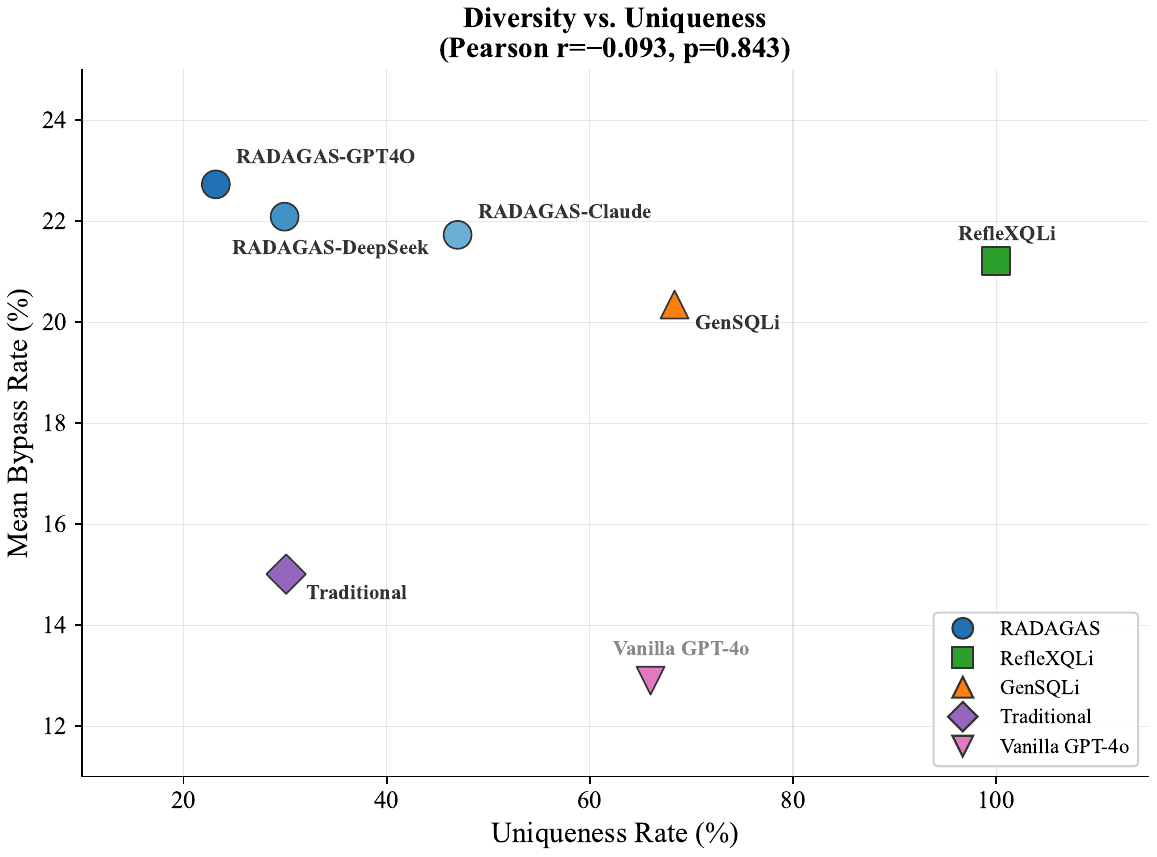}
\caption{Uniqueness rate vs.\ WAF bypass rate across all seven systems ($r=-0.093$, $p=0.843$, not significant). RADAGAS-GPT4o achieves the highest bypass (22.73\%) with only 23.18\% uniqueness, while RefleXQLi (100\% unique) ranks fourth (21.21\%), demonstrating that uniqueness alone does not determine bypass effectiveness.}
\label{fig:diversity_paradox}
\end{figure}

\textbf{Rule based WAF resilience:} Interestingly, rule based WAFs show very strong prevention rate especially with higher paranoia levels with PL2 and PL3. Even in lowest paranoia level PL1, GenSQLi performed highest bypass level with 16.26\%. However, qualitative analysis showed that GenSQLi's 813 ModSecurity PL1 bypass payloads reveal that a significant portion are false positives with 11.2\% that are malformed artifacts and 29.3\% are tautological expressions which do not expose real threat. We manually classified all 813 bypass payloads into four categories:
\begin{itemize}
    \item data exfiltration queries like (UNION SELECT)
    \item tautological expressions like (OR 1=1, AND ''='')
    \item time-based blind (SLEEP, BENCHMARK)
    \item malformed SQL
\end{itemize}
Additionally, the experiment did not show a significant correlation ($r=-0.093$, $p=0.843$) in between bypass rate and uniqueness as in the Fig.~\ref{fig:diversity_paradox}. If the bypass rate is higher in certain patterns, the systems may achieve high bypass rate with less unique payloads. However this creates a practical weakness where dynamic WAFs can prevent certain patterns and the generator needs to cover more diverse pattern to sustain bypass capability.

\textbf{MySQL execution:} RADAGAS variants generated highest valid execution rate with 60--64\%. This is followed by GenSQLi with 36.96\% and RefleXQLi (27.80\%). Traditional and Vanilla GPT-4o stayed at (23.12\%) and (21.44\%) respectively. Since RAG systems use curated knowledgebase, the execution rate is higher. Whereas GenSQLi generated tautologies which are similar simple payloads and passed the execution test. Traditional and zero-shot systems could not achieve that high rates.

\textbf{Commercial WAF performance:} RADAGAS-GPT4o performed the best Cloudflare bypass rate with 49.50\%. Our another novel algorithm RefleXQLi achieved 33.00\% bypass rate on AWS WAF. Whereas, GenSQLi performed 20.58\% on Cloudflare. These results show that commercial WAF providers adopted heterogeneous detection strategies, since different generation systems bypass different rule gaps.

\textbf{Complementary strengths:}
Results revealed that the success rate is system specific and each system has different gaps and patterns. The highest bypass rates for WAF-Brain is 92.49\% (RADAGAS-DeepSeek), CNN-WAF is 80.48\% (RADAGAS-Claude), Cloudflare is 49.50\% (GPT-4o) and rule based WAFs are dominated by GenSQLi for Coraza PL1 with 14.74\% and ModSecurity PL1 with 16.26\%. This shows that combination of different generation strategies can maximize the coverage.

\FloatBarrier
\subsection{Seed Effect Analysis}
\label{subsec:seed_effect}

This section presents the seed effect ablation experiment for RefleXQLi and GenSQLi. In the previous sections, we investigated the effect of configuration parameters. However there is still a big question on effectiveness of seed mechanism. To answer this question, we created a controlled seed effect experiment comparing two systems that use the same GenSQLi V2 paper seeds~\cite{babaey2025gensqli} and denoted these system variants as GenSQLi V2 and RefleXQLi V2.

\subsubsection{System Comparison}

For the experiment, we prepared 11 GenSQLi seeds from the paper and converted them to string context with quote break prefix and comment suffix, since our backend services are vulnerable to the parameter in this format. Then we generated 5$\times$1,000 payloads for each of systems.
\begin{figure*}[!t]
\centering
\includegraphics[width=0.85\textwidth]{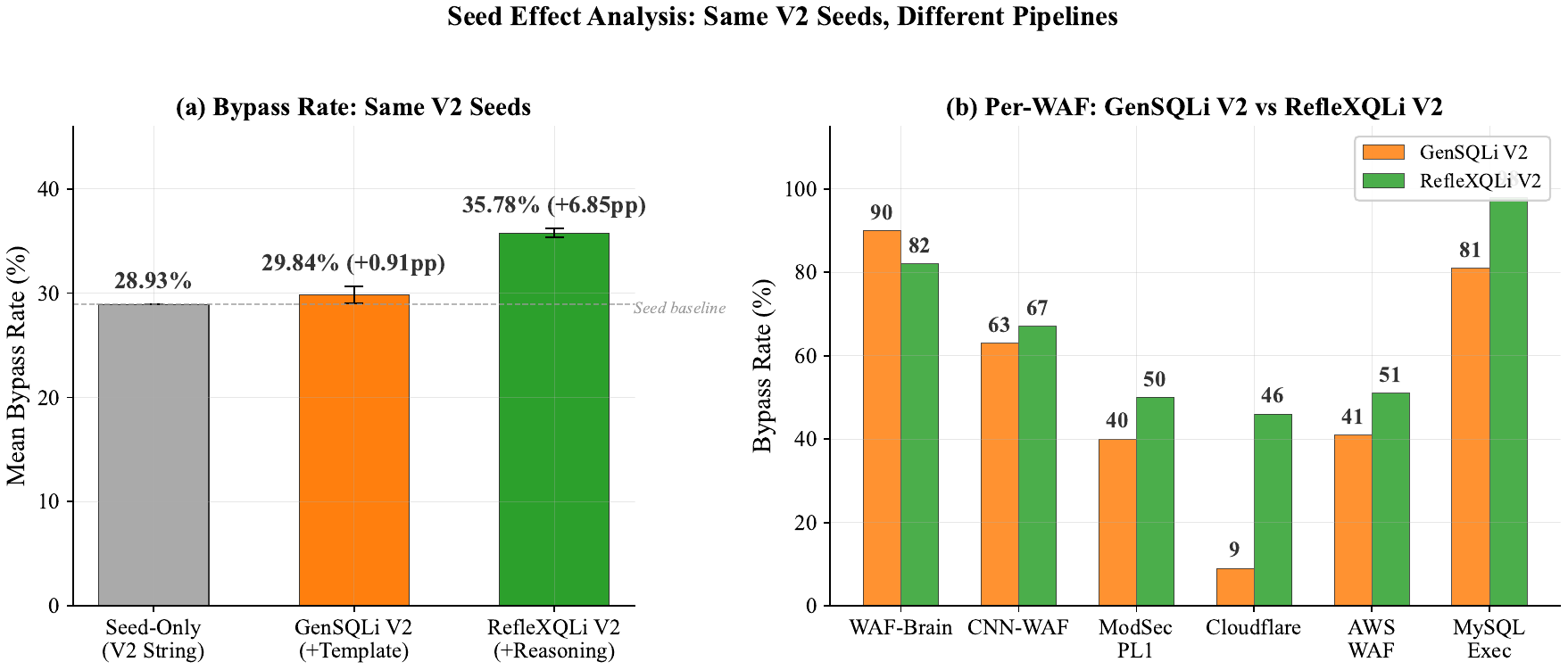}
\caption{Seed effect analysis. Left: Overall bypass rate comparison showing raw V2 seeds (28.93\%), GenSQLi V2 (+0.91pp), and RefleXQLi V2 (+6.85pp). CoT reasoning (RefleXQLi provides 7.5$\times$ more improvement than template-based mutation (GenSQLi). Right: Per-WAF comparison between GenSQLi V2 and RefleXQLi V2 using the same seeds}
\label{fig:seed_effect}
\end{figure*}
\begin{table}[t]
\caption{Seed Effect: Seed-Only bypass rate represents the raw paper seeds tested directly against all 11 WAFs without any LLM processing.}
\label{tab:seed_effect_bypass}
\begin{tabular*}{\columnwidth}{@{\extracolsep\fill} lcccc @{}}
\toprule
\textbf{Configuration} & \textbf{Bypass (\%)} & \textbf{$\sigma$ (\%)} & \textbf{MySQL (\%)} & \textbf{$\Delta$ vs Seed} \\
\midrule
Seed-Only (V2 string) & 28.93 & --- & --- & --- \\
GenSQLi V2 & 29.84 & 0.79 & 80.97 & +0.91 \\
RefleXQLi V2 & \textbf{35.78} & \textbf{0.44} & \textbf{98.36} & \textbf{+6.85} \\
\bottomrule
\end{tabular*}
\end{table}

Fig.~\ref{fig:seed_effect} visualizes the seed effect comparison, showing both bypass improvement of the seed and v2 models at left and breakdown of the performance per WAF at right. Table~\ref{tab:seed_effect_bypass} shows a significant contrast between two V2 pipelines. While GenSQLi V2 adds +0.91pp performance with 29.84\% bypass rate over original seed bypass rate (28.93\%), RefleXQLi V2 achieves +6.85pp and reaches 35.78\% bypass rate demonstrating that template based mutation techniques provide minimal improvement on raw seeds, while CoT based reasoning techniques achieve particularly significant improvement on the overall bypass rate. Valid MySQL execution rate further differentiates the systems where RefleXQLi V2 achieves 98.36\% and GenSQLi V2 has 80.97\% semantic correctness showing CoT produces both semantically and syntactically valid generation outputs rather than template based approaches.

\subsubsection{Seed Effect Comparison}

\begin{table}[t]
\caption{Per-WAF Bypass Rates with Same V2 Seeds}
\label{tab:seed_effect_waf}
\begin{tabular*}{\columnwidth}{@{\extracolsep\fill} lccc @{}}
\toprule
\textbf{WAF} & \textbf{GenSQLi V2} & \textbf{RefleXQLi V2} & \textbf{$\Delta$} \\
& \textbf{(\%)} & \textbf{(\%)} & \\
\midrule
WAF-Brain & \textbf{89.71} & 81.83 & $-$7.88 \\
CNN-WAF & 62.80 & \textbf{66.55} & +3.75 \\
ModSec PL1 & 39.80 & \textbf{50.19} & +10.39 \\
ModSec PL2 & \textbf{0.96} & 0.00 & $-$0.96 \\
ModSec PL3 & \textbf{0.96} & 0.00 & $-$0.96 \\
Cloudflare & 8.87 & \textbf{45.96} & +37.09 \\
AWS WAF & 40.79 & \textbf{50.65} & +9.86 \\
MySQL Exec & 80.97 & \textbf{98.36} & +17.39 \\
Coraza PL1 & \textbf{1.22} & 0.02 & $-$1.20 \\
Coraza PL2 & \textbf{1.09} & 0.00 & $-$1.09 \\
Coraza PL3 & \textbf{1.09} & 0.00 & $-$1.09 \\
\midrule
\textbf{Mean} & 29.84 & \textbf{35.78} & +5.94 \\
\bottomrule
\end{tabular*}
\end{table}
Table~\ref{tab:seed_effect_waf} shows granular performance for two systems using V2 seeds per WAF which reveals the complementary strengths of both systems. The results reveal that RefleXQLi V2 shows higher performance than GenSQLi V2 on commercial WAFs with Cloudflare (+37.09pp) and AWS WAF (+9.86pp) suggesting that CoT reasoning discovers obfuscation techniques effectively on the commercial WAFs. This pattern is also similar in rule based WAF ModSecurity PL1 (+10.39pp) revealing the effectiveness of CoT over in-context learning generation. However, GenSQLi V2 shows better performance on WAF-Brain with +7.88pp which suggests that usage of both systems provides a complementary coverage on overall systems.

\subsubsection{Diversity Comparison Under Same Seeds}

\begin{table}[t]
\caption{Seed Effect Diversity: Same Seeds, Different Pipelines}
\label{tab:seed_effect_diversity}
\begin{tabular*}{\columnwidth}{@{\extracolsep\fill} lcc @{}}
\toprule
\textbf{Metric} & \textbf{GenSQLi V2} & \textbf{RefleXQLi V2} \\
\midrule
Uniqueness (\%) & 66.14 & \textbf{95.84} \\
Semantic & 0.3787 & \textbf{0.4385} \\
Lexical & 0.6097 & \textbf{0.6312} \\
Contextual & \textbf{0.1265} & 0.1257 \\
N-gram & 0.7527 & \textbf{0.8176} \\
AST & 0.6814 & \textbf{0.6941} \\
Functional & 0.3830 & \textbf{0.3999} \\
\midrule
Total Diversity & 0.4887 & \textbf{0.5178} \\
\bottomrule
\end{tabular*}
\end{table}
Table~\ref{tab:seed_effect_diversity} shows the comparison of six post-hoc diversity metric performance of two V2 systems. The results reveal that RefleXQLi V2 has higher diversity levels for five out of six metrics with 0.5178 vs.\ 0.4887 overall diversity score and substantially higher uniqueness rate with 95.84\% vs.\ 66.14\%. GenSQLi V2 outperforms RefleXQLi V2 for only contextual diversity (BERTScore) with 0.1265 vs.\ 0.1257 which is a negligible difference. This result demonstrates that CoT based reasoning provides higher diversity and produces more effective payloads than in-context learning generation.

\subsubsection{Seed Quality and LLM Reasoning}

There are two significant findings of the seed effect analysis experiment. The first and most significant result suggests that the seed quality provides a strong baseline where the raw seeds achieved 28.93\% success rate without any LLM intervention. The second finding is RefleXQLi V2 has a higher MySQL execution rate with 98.36\% while GenSQLi has 80.97\% levels suggesting that CoT reasoning is very effective in creating syntactically correct outputs. Overall results show that investment in both seed engineering and CoT reasoning yields successful returns for adversarial generation.

\section{Discussion}
\label{sec:discussion}

\subsection{Role of Diversity in Adversarial Generation}

The multi-method analysis across three correlation methods Pearson, Spearman, Kendall shows that out of six diversity metrics there is no significant correlation with bypass rate of the systems. The only significant finding is, negative Spearman correlation of N-gram diversity suggests that lower structural lexical diversity contributes higher bypass rates.

Intra system analysis with 48 configurations for RADAGAS variants shows significant positive correlation between diversity threshold and bypass rate with $r=+0.371$, $p=0.009$. That means, in a single setup and architecture, tuning diversity improves performance higher than architecturally different systems. That suggests that architectural design, retrieval strategy and reasoning mechanisms dominate over diversity metrics and that makes diversity effect negligible.

\subsection{Model Specific Behavior and Training Methodology}

Out of 48 configurations, three foundation models of RADAGAS showed optimal performance under different $T$ values with GPT-4o at $T=0.1$, DeepSeek at $T=0.9$, Claude at $T=0.6$ suggesting that $T$ values are model specific.

These findings reveal direct implications for adversarial generation. There is no generic parameter set that can be transferred across models to obtain optimal efficiency for adversarial generation, since each foundation model requires independent empirical tuning.

\subsection{Architectural Tradeoffs}

Our evaluation shows complementary strengths across different architectural designs:

RAG based systems (RADAGAS) achieve highest performance across seven generation systems. The success lies in leveraging proven successful payloads from curated lists. RAG based systems had a very poor performance against rule based WAFs where the established rules cover the known payloads very well. However, RAG based systems were successful against AI/ML based WAFs where DeepSeek showed 92.49\% bypass rate on WAF-Brain suggesting ML based WAFs were not trained well to protect against known payloads.

CoT+Adversarial (RefleXQLi) performed a competitive performance with 21.21\%, $\sigma=0.37\%$ in V1 mode. In V2 mode, seed effect ablation (Section~\ref{subsec:seed_effect}) shows adding high quality seeds increases the bypass rate to 35.78\% with 98.36\% MySQL execution suggesting that seed quality and CoT reasoning have big impact on adversarial generation.

Baseline systems perform lower. In-context learning algorithm GenSQLi achieves highest nominal bypass rates on rule based WAFs, with majority of payloads being tautological expressions. Vanilla GPT-4o and Traditional SQLi showed poor performance compared to other systems with 12.90\%, 15.01\% bypass rates respectively.

Overall results show that, complementary WAF results suggests, combining CoT, RAG and In context learning generators together, maximizes the coverage across diverse WAFs.

\subsection{Technical Limitations and Considerations}
There are five main aspects in terms of technical limitations.

\textbf{WAF coverage:} While our experiment has a rich set of WAFs with commercial, rule based and AI/ML based, the results may not be generalized to all commercial WAFs or custom enterprise settings and deployments~\cite{Razzaq2013WAFSurvey}. Our tests were performed against specific WAF versions in years 2025 and 2026.

\textbf{Attack type scope:} We limited the attack scope with SQL Injection. However, generalization to other attack types (i.e XSS, Command Injection, SSRF) may require another empirical study with different approach.

\textbf{Database specificity:} In our experiments, we used MySQL 8.0 to check SQLi validity. However, SQL dialects have differences across PostgreSQL, MSSQL and Oracle that may affect structural and semantic correctness results~\cite{Clarke2012SQLBook}.

\textbf{Sample size for cross system correlation:} The cross system correlation analysis is done with $N=7$ data points limiting statistical power and preventing us from definitive conclusions about the correlation. To mitigate this gap, we employed three complementary strategies. First, for small samples, we employed non parametric methods (Spearman, Kendall $\tau$). Second, to quantify uncertainty, we computed bootstrap 95\% confidence intervals with 10,000 resamples and lastly, we validated the findings with intra-system analysis with 48 RADAGAS configurations ($N=48$) which provided sufficient statistical power ($\beta > 0.80$).

\textbf{Temporal validity:} Since the systems evolve continuously, our results reflect the state of the LLM and WAF detection systems as of early 2026.

\subsection{Future Directions}

As a next step of this study, we have several research directions from our findings. Since our tests were only SQL injection specific, verifying the results with other attack types (XSS, SSRF, Command Injection etc.) will extend the results to provide more generalized context. Another direction can be extending the coverage on other foundational models to analyze model specific optimal parameter settings and patterns more. Moreover, extending execution validation within multiple databases to provide more comprehensive landscape for all SQL injection types to generate DBMS specific attack generation strategies.

\subsection{Ethical Considerations and Responsible Disclosure}
\label{subsec:ethics}

All experiments were conducted in isolated environments. The generated SQLi payloads were executed in self hosted environment with necessary authorizations. We tested Commercial WAFs with our own private testing accounts backed by specially crafted dummy web endpoints to measure the effectiveness of the rules. The generated payloads have never been used against unauthorized systems. The source code is kept as private to prevent any potential misuse of the system to generate malicious payloads.

\section{Conclusion}
\label{sec:conclusion}

This paper presents an extensive evaluation of pure LLM based SQL Injection attack generation methods and their effectiveness through two novel adversarial generation systems RADAGAS (three variants, RAG based) and RefleXQLi (CoT based) benchmarked alongside three baseline systems. These seven systems were tested against ten WAFs and an execution test. The overall experiment generated 240,000 payloads and performed 2.2 million WAF tests constituting one of the largest benchmarking studies of LLM based adversarial SQL injection generation.

Our experiments yielded findings that have theoretical and practical significance. While the expectation was that diversity drives the bypass effectiveness, the results show that the contribution of diversity is minimal or negligible considering the huge effect of seed effect combined with CoT reasoning. Practically, our results provide solid real life deployment guidelines where complementary usage of different systems has more coverage: RAG systems for AI/ML WAFs, CoT for stable adversarial generation, and seed driven CoT provides the highest bypass rate.

\backmatter

\section*{AI Transparency}
In this paper, generative AI methods are compared in a scientific framework. Apart from that, AI-based tools are utilized for grammar checks and corrections following the human-in-the-loop principle, i.e., all AI suggestions were carefully checked. The authors reviewed and edited all AI-assisted improvements and take full responsibility for the final manuscript.

\section*{Declarations}

\begin{itemize}
\item \textbf{Competing interests:} The authors have no relevant financial or non-financial interests to disclose.
\item \textbf{Data availability:} Data will be made available upon reasonable request.
\item \textbf{Funding:} Not applicable
\item \textbf{Conflict of interest:} The authors declare that they have no conflict of interest.
\item \textbf{Ethics approval and consent to participate:} Not applicable
\item \textbf{Consent for publication:} Not applicable
\item \textbf{Author contribution:} A. Karakoc: Conceptualization, Methodology, Software, Investigation, Writing - Original Draft, Visualization. H. B. Yilmaz: Supervision, Conceptualization, Writing - Review \& Editing.
\end{itemize}

\bibliography{references}

@misc{owasp2021,
  author = {{OWASP Foundation}},
  title = {{OWASP} Top 10-2021: The Ten Most Critical Web Application Security Risks},
  year = {2021},
  url = {https://owasp.org/Top10/}
}

@misc{sqlmap2006,
  author = {Damele, Bernardo and Stampar, Miroslav},
  title = {{SQLMAP}: Automatic {SQL} Injection and Database Takeover Tool},
  year = {2006},
  url = {https://sqlmap.org/},
  note = {Open-source penetration testing tool}
}

@inproceedings{Kindy2011SQLi,
  author = {Kindy, Daizong Ali and Pathan, Al-Sakib Khan},
  title = {A Survey on {SQL} Injection: Vulnerabilities, Attacks, and Prevention Techniques},
  booktitle = {IEEE International Symposium on Consumer Electronics (ISCE)},
  year = {2011},
  pages = {468-471},
  doi = {10.1109/ISCI.2011.5973873}
}

@book{Clarke2012SQLBook,
  author = {Clarke, Justin},
  title = {{SQL} Injection Attacks and Defense},
  publisher = {Elsevier},
  year = {2012},
  address   = {Waltham, MA}
}

@Article{babaey2025gensqli,
    AUTHOR = {Babaey, Vahid and Ravindran, Arun},
    TITLE = {{GenSQLi}: A Generative Artificial Intelligence Framework for Automatically Securing Web Application Firewalls Against Structured Query Language Injection Attacks},
    JOURNAL = {Future Internet},
    VOLUME = {17},
    YEAR = {2025},
    NUMBER = {1},
    ARTICLE-NUMBER = {8},
    DOI = {10.3390/fi17010008}
}

@inproceedings {deng2024pentestgpt,
    author = {Gelei Deng and Yi Liu and Victor Mayoral-Vilches and Peng Liu and Yuekang Li and Yuan Xu and Tianwei Zhang and Yang Liu and Martin Pinzger and Stefan Rass},
    title = {{PentestGPT}: Evaluating and Harnessing Large Language Models for Automated Penetration Testing},
    booktitle = {USENIX Security Symposium (USENIX Security 24)},
    year = {2024},
    isbn = {978-1-939133-44-1},
    pages = {847--864},
    url = {https://www.usenix.org/conference/usenixsecurity24/presentation/deng}
}

@misc{fang2025llm4vuln,
      author={Yuqiang Sun and Daoyuan Wu and Yue Xue and Han Liu and Wei Ma and Lyuye Zhang and Yang Liu and Yingjiu Li},
      title={{LLM4Vuln}: A Unified Evaluation Framework for Decoupling and Enhancing {LLMs}' Vulnerability Reasoning},   
      year={2025},
      eprint={2401.16185},
      archivePrefix={arXiv},
      url={https://arxiv.org/abs/2401.16185}, 
}

@misc{liu2025promptinject,
      author={Yi Liu and Gelei Deng and Yuekang Li and Kailong Wang and Zihao Wang and Xiaofeng Wang and Tianwei Zhang and Yepang Liu and Haoyu Wang and Yan Zheng and Leo Yu Zhang and Yang Liu}, 
      title={Prompt Injection attack against {LLM}-integrated Applications}, 
      year={2025},
      eprint={2306.05499},
      archivePrefix={arXiv},
      url={https://arxiv.org/abs/2306.05499}, 
}

@Article{sqligpt,
AUTHOR = {Gui, Zhiwen and Wang, Enze and Deng, Binbin and Zhang, Mingyuan and Chen, Yitao and Wei, Shengfei and Xie, Wei and Wang, Baosheng},
TITLE = {{SqliGPT}: Evaluating and Utilizing Large Language Models for Automated {SQL} Injection Black-Box Detection},
JOURNAL = {Applied Sciences},
VOLUME = {14},
YEAR = {2024},
NUMBER = {16},
ARTICLE-NUMBER = {6929},
DOI = {10.3390/app14166929}
}

@inproceedings{llmsqli,
  author={Yang, Tinghui and Jiang, Zhiyuan and Wang, Yongjun},
  title={{LLMSQLi}: A black-box web {SQLi} detection tool based on large language model},
  booktitle={IEEE International Conference on Big Data \& Artificial Intelligence \& Software Engineering (ICBASE)},
  pages={629--633},
  year={2024}
}

@misc{openai2024gpt4,
      title={{GPT-4 Technical Report}}, 
      author={OpenAI},
      year={2024},
      archivePrefix={arXiv},
      url={https://arxiv.org/abs/2303.08774}, 
}

@article{anthropic2024claude,
  title={The claude 3 model family: Opus, sonnet, haiku},
  author={Anthropic-AI},
  journal={Claude-3 Model Card},
  volume={1},
  number={1},
  pages={4},
  year={2024},
  url = {https://www.anthropic.com/claude}
}

@misc{deepseek2024coder,
      author={DeepSeek-AI},
      title={DeepSeek-Coder-V2: Breaking the Barrier of Closed-Source Models in Code Intelligence}, 
      year={2024},
      archivePrefix={arXiv},
      url={https://arxiv.org/abs/2406.11931}, 
}

@inproceedings{carbonell1998mmr,
    author = {Carbonell, Jaime and Goldstein, Jade},
    title = {{The use of MMR, diversity-based reranking for reordering documents and producing summaries}},
    year = {1998},
    publisher = {Association for Computing Machinery},
    address = {New York, NY, USA},
    doi = {10.1145/290941.291025},
    booktitle = {Proceedings of the  Annual International ACM SIGIR Conference on Research and Development in Information Retrieval},
    pages = {335–336}
}

@inproceedings{zhang2024temperature,
    author = "Renze, Matthew",
    title = "The Effect of Sampling Temperature on Problem Solving in Large Language Models",
    booktitle = "Findings of the Association for Computational Linguistics (EMNLP)",
    month = nov,
    year = "2024",
    address = "Miami, Florida, USA",
    publisher = "Association for Computational Linguistics",
    doi = "10.18653/v1/2024.findings-emnlp.432",
    pages = "7346--7356",
}

@inproceedings{wei2022chain,
 author = {Wei, Jason and Wang, Xuezhi and Schuurmans, Dale and Bosma, Maarten and ichter, brian and Xia, Fei and Chi, Ed and Le, Quoc V and Zhou, Denny},
 title = {{Chain-of-Thought Prompting Elicits Reasoning in Large Language Models}},
 booktitle = {Advances in Neural Information Processing Systems},
 pages = {24824--24837},
 url = {https://proceedings.neurips.cc/paper_files/paper/2022/file/9d5609613524ecf4f15af0f7b31abca4-Paper-Conference.pdf},
 volume = {35},
 year = {2022}
}

@inproceedings{brown2020language,
 author = {Brown, Tom and Mann, Benjamin and Ryder, Nick and Subbiah, Melanie and Kaplan, Jared D and Dhariwal, Prafulla and Neelakantan, Arvind and Shyam, Pranav and Sastry, Girish and Askell, Amanda and Agarwal, Sandhini and Herbert-Voss, Ariel and Krueger, Gretchen and Henighan, Tom and Child, Rewon and Ramesh, Aditya and Ziegler, Daniel and Wu, Jeffrey and Winter, Clemens and Hesse, Chris and Chen, Mark and Sigler, Eric and Litwin, Mateusz and Gray, Scott and Chess, Benjamin and Clark, Jack and Berner, Christopher and McCandlish, Sam and Radford, Alec and Sutskever, Ilya and Amodei, Dario},
 title = {{Language Models are Few-Shot Learners}},
 booktitle = {Advances in Neural Information Processing Systems},
 pages = {1877--1901},
 url = {https://proceedings.neurips.cc/paper_files/paper/2020/file/1457c0d6bfcb4967418bfb8ac142f64a-Paper.pdf},
 volume = {33},
 year = {2020}
}

@inproceedings{Dong2024Survey,
    title = "A Survey on In-context Learning",
    author = "Dong, Qingxiu  and
      Li, Lei  and
      Dai, Damai  and
      Zheng, Ce  and
      Ma, Jingyuan  and
      Li, Rui  and
      Xia, Heming  and
      Xu, Jingjing  and
      Wu, Zhiyong  and
      Chang, Baobao  and
      Sun, Xu  and
      Li, Lei  and
      Sui, Zhifang",
    booktitle = "Proceedings of the  Conference on Empirical Methods in Natural Language Processing",
    year = "2024",
    address = "Miami, Florida, USA",
    publisher = "Association for Computational Linguistics",
    doi = "10.18653/v1/2024.emnlp-main.64",
    pages = "1107--1128",
}

@inproceedings{Goodfellow2014GAN,
 author = {Goodfellow, Ian J. and Pouget-Abadie, Jean and Mirza, Mehdi and Xu, Bing and Warde-Farley, David and Ozair, Sherjil and Courville, Aaron and Bengio, Yoshua},
 title = {Generative Adversarial Nets},
 booktitle = {Advances in Neural Information Processing Systems},
 url = {https://proceedings.neurips.cc/paper_files/paper/2014/file/f033ed80deb0234979a61f95710dbe25-Paper.pdf},
 volume = {27},
 year = {2014}
}

@misc{Holtzman2020Nucleus,
      author={Ari Holtzman and Jan Buys and Li Du and Maxwell Forbes and Yejin Choi},
      title={The Curious Case of Neural Text Degeneration}, 
      year={2020},
      archivePrefix={arXiv},
      url={https://arxiv.org/abs/1904.09751}, 
}

@article{Meister2023Locally,
    author = "Meister, Clara  and
      Pimentel, Tiago  and
      Wiher, Gian  and
      Cotterell, Ryan",
    title = "Locally Typical Sampling",
    journal = "Transactions of the Association for Computational Linguistics",
    volume = "11",
    year = "2023",
    address = "Cambridge, MA",
    publisher = "MIT Press",
    doi = "10.1162/tacl_a_00536",
    pages = "102--121"
}

@misc{zhang2020bertscore,
      author={Tianyi Zhang and Varsha Kishore and Felix Wu and Kilian Q. Weinberger and Yoav Artzi},
      title={{BERTScore: Evaluating Text Generation with BERT}},    
      year={2020},
      archivePrefix={arXiv},
      url={https://arxiv.org/abs/1904.09675}, 
}

@inproceedings{levenshtein1966binary,
  title={Binary codes capable of correcting deletions, insertions, and reversals},
  author={Levenshtein, Vladimir I},
  booktitle={Soviet Physics Doklady},
  volume={10},
  number={8},
  pages={707--710},
  year={1966},
  organization={Soviet Union}
}

@inproceedings{reimers2019sentence,
    author = "Reimers, Nils  and
      Gurevych, Iryna",
    title = {{Sentence-{BERT}: Sentence Embeddings using {S}iamese {BERT}-Networks}},
    booktitle = "Proceedings of the  Conference on Empirical Methods in Natural Language Processing and the  International Joint Conference on Natural Language Processing (EMNLP-IJCNLP)",
    year = "2019",
    address = "Hong Kong, China",
    doi = "10.18653/v1/D19-1410",
    pages = "3982--3992"
}

@article{johnson2019billion,
  author={Johnson, Jeff and Douze, Matthijs and J{\'e}gou, Herv{\'e}},
  title={{Billion-scale similarity search with GPUs}},  
  journal={IEEE Transactions on Big Data},
  volume={7},
  number={3},
  pages={535--547},
  year={2019}
}

@inproceedings{muennighoff2023mteb,
    author = "Muennighoff, Niklas  and
          Tazi, Nouamane  and
          Magne, Loic  and
          Reimers, Nils",
    title = {{MTEB: Massive Text Embedding Benchmark}},
    booktitle = "Proceedings of the  Conference of the European Chapter of the Association for Computational Linguistics",
    year = "2023",
    address = "Dubrovnik, Croatia",
    doi = "10.18653/v1/2023.eacl-main.148",
    pages = "2014--2037",
}

@article{Manes2019FuzzingSurvey,
  author={Manes, Valentin JM and Han, HyungSeok and Han, Choongwoo and Cha, Sang Kil and Egele, Manuel and Schwartz, Edward J and Woo, Maverick},
  title={{The art, science, and engineering of fuzzing: A survey}},
  journal={IEEE Transactions on Software Engineering},
  volume={47},
  number={11},
  pages={2312--2331},
  year={2019}
}

@misc{Zalewski2017AFL,
  title={American fuzzy lop (AFL) fuzzer},
  author={Zalewski, Michal},
  url={https://lcamtuf.coredump.cx/afl},
  year={2017}
}

@inproceedings{Bohme2016Coverage,
    author = {B\"{o}hme, Marcel and Pham, Van-Thuan and Roychoudhury, Abhik},
    title = {Coverage-based Greybox Fuzzing as Markov Chain},
    year = {2016},
    address = {New York, NY, USA},
    doi = {10.1145/2976749.2978428},
    booktitle = {Proceedings of the ACM SIGSAC Conference on Computer and Communications Security},
    pages = {1032–1043},
    location = {Vienna, Austria}
}

@inproceedings{Lemieux2018FairFuzz,
    author = {Lemieux, Caroline and Sen, Koushik},
    title = {FairFuzz: a targeted mutation strategy for increasing greybox fuzz testing coverage},
    year = {2018},
    address = {New York, NY, USA},
    doi = {10.1145/3238147.3238176},
    booktitle = {Proceedings of the  ACM/IEEE International Conference on Automated Software Engineering},
    pages = {475–485},
    location = {Montpellier, France}
}

@inproceedings{FuzzQL2020,
    author = {Zhong, Rui and Chen, Yongheng and Hu, Hong and Zhang, Hangfan and Lee, Wenke and Wu, Dinghao},
    title = {{SQUIRREL: Testing Database Management Systems with Language Validity and Coverage Feedback}},
    year = {2020},
    address = {New York, NY, USA},
    doi = {10.1145/3372297.3417260},
    booktitle = {Proceedings of the ACM SIGSAC Conference on Computer and Communications Security},
    pages = {955–970},
    location = {Virtual Event, USA}
}

@inproceedings {SQIRL,
author = {Salim Al Wahaibi and Myles Foley and Sergio Maffeis},
title = {{SQIRL}: {Grey-Box} Detection of {SQL} Injection Vulnerabilities Using Reinforcement Learning},
booktitle = {USENIX Security Symposium (USENIX Security)},
year = {2023},
address = {Anaheim, CA},
pages = {6097--6114},
url = {https://www.usenix.org/conference/usenixsecurity23/presentation/al-wahaibi},
publisher = {USENIX Association}
}

@inproceedings{Hu2020RL,
  title={Automated penetration testing using deep reinforcement learning},
  author={Hu, Zhenguo and Beuran, Razvan and Tan, Yasuo},
  booktitle={IEEE European Symposium on Security and Privacy Workshops (EuroS\&PW)},
  pages={2--10},
  year={2020}
}

@misc{modsecurity2024,
  author = {OWASP},
  title = {{ModSecurity: Open Source Web Application Firewall}},
  year = {2024},
  url = {https://modsecurity.org/}
}

@misc{Coraza2024,
  author = {OWASP},
  title = {{Coraza Web Application Firewall}},
  year = {2024},
  url = {https://coraza.io/}
}

@article{Razzaq2013WAFSurvey,
author = {Abdul Razzaq and Khalid Latif and H. Farooq Ahmad and Ali Hur and Zahid Anwar and Peter Charles Bloodsworth},
title = {Semantic security against web application attacks},
journal = {Information Sciences},
volume = {254},
pages = {19-38},
year = {2014},
doi = {https://doi.org/10.1016/j.ins.2013.08.007}
}

@inproceedings{appelt2015websec,
  author={Appelt, Dennis and Nguyen, Cu D and Briand, Lionel},
  title={{Behind an application firewall, are we safe from SQL injection attacks?}},
  booktitle={IEEE International Conference on Software Testing, Verification and Validation (ICST)},
  pages={1--10},
  year={2015}
}

@misc{WAFBrain2018,
  author = {{BBVA-Labs}},
  title = {{WAF-Brain: Machine Learning Based Web Application Firewall}},
  year = {2018},
  url = {https://github.com/BBVA/waf-brain}
}

@article{CNNWAF2019,
  author={Pan, Yao and Sun, Fangzhou and Teng, Zhongwei and White, Jules and Schmidt, Douglas C and Staples, Jacob and Krause, Lee},
  title={Detecting web attacks with end-to-end deep learning}, 
  journal={Journal of Internet Services and Applications},
  volume={10},
  number={1},
  pages={1--22},
  year={2019},
  publisher={Springer}
}

@article{Pearson1920Correlation,
 author = {Karl Pearson},
 URL = {http://www.jstor.org/stable/2331722},
 journal = {Biometrika},
 number = {1},
 pages = {25--45},
 title = {{Notes on the History of Correlation}},
 volume = {13},
 year = {1920}
}

@article{Fisher1915Correlation,
 author = {Fisher, Ronald A.},
 title = {Frequency Distribution of the Values of the Correlation Coefficient in Samples from an Indefinitely Large Population},
 URL = {http://www.jstor.org/stable/2331838},
 journal = {Biometrika},
 number = {4},
 pages = {507--521},
 volume = {10},
 year = {1915}
}

@book{cohen2013statistical,
  title={Statistical power analysis for the behavioral sciences},
  author={Cohen, Jacob},
  year={2013},
  publisher = {Routledge},
  address   = {New York}
}

@misc{PortSwigger2024,
  author = {{PortSwigger}},
  title = {{Web Security Academy: SQL Injection}},
  year = {2024},
  url = {https://portswigger.net/web-security/sql-injection}
}

@misc{PayloadsAllTheThings,
  author = {{swisskyrepo} and contributors},
  title = {{PayloadsAllTheThings}: A List of Useful Payloads and Bypass for Web Application Security},
  year = {2024},
  url = {https://github.com/swisskyrepo/PayloadsAllTheThings},
  note = {Community-maintained security payload repository}
}

\end{document}